\documentclass[aps,prb,onecolumn,floats,epsfig,pdflatex]{revtex4}
\usepackage{color,ulem,verbatim,bbold,float,wrapfig}
\usepackage{amssymb,amsbsy,amsmath,mathrsfs}
\usepackage{soul}
\usepackage{epsfig,subfigure}

\usepackage[colorlinks,linktocpage,bookmarks=false,citecolor=blue,linkcolor=red,urlcolor=blue]{hyperref}

\newcommand{\beq}{\begin{equation}}
\newcommand{\eeq}{\end{equation}}
\newcommand{\bea}{\begin{eqnarray}}
\newcommand{\eea}{\end{eqnarray}}

\begin{document}

\title{Phases and dynamics of ultracold bosons in a tilted optical lattice}

\author{K. Sengupta}

\affiliation{School of Physical Sciences, Indian Association for the
Cultivation of Science, 2A and 2B Raja S. C. Mullick Road, Jadavpur
700032, India. }

\date{\today}

\begin{abstract}

We present a brief overview of the phases and dynamics of ultracold
bosons in an optical lattice in the presence of a tilt. We begin
with a brief summary of the possible experimental setup for
generating the tilt. This is followed by a discussion of the
effective low-energy model for these systems and its equilibrium
phases. We also chart the relation of this model to the recently
studied system of ultracold Rydberg atoms. Next, we discuss the
non-equilibrium dynamics of this model for quench, ramp and periodic
protocols with emphasis on the periodic drive which can be
understood in terms of an analytic, albeit perturbative, Floquet
Hamiltonian derived using Floquet perturbation theory (FPT).
Finally, taking cue from the Floquet Hamiltonian of the periodically
driven tilted boson chain, we discuss a spin model which exhibits
Hilbert space fragmentation and exact dynamical freezing for wide
range of initial states.

\end{abstract}

\maketitle

\section{Introduction}
\label{sec:int}

The physics of ultracold bosonic atoms in an optical lattice has
attracted tremendous attention in recent years \cite{bref1,
bref1a,bref2,bref3,brefrev}. This enthusiasm stemmed from the fact
that such a boson system acts as one of the simplest emulator of the
Bose-Hubbard model \cite{bref5}. Thus the study of the low-energy
physics of these bosons allows one to access the
superfluid-insulator quantum phase transition for ultracold bosons
which is well-known to be present in the phase diagram of a clean
Bose-Hubbard model
\cite{fisher1,nandini1,tvr1,free1,nicolas1,tref1}.

The Mott phase of the bosons in this system, at integer filling,
constitutes a localized state with no broken symmetry. It was long
realized that additional broken translational or other discrete
symmetries may lead to interesting strongly-interacting phases of
matter within the Mott phase. To this end, several theoretical
suggestions have been put forth. These include study of bosons with
nearest-neighbor repulsive interaction leading to the possibility of
different competing density-wave ground states at half-filling due
to broken translational symmetry. The precise nature of these states
depends on the geometry of the underlying lattice leading to a
projective symmetry group (PSG) based classification of the possible
Mott phases \cite{balents1,balents2,yb1}. In addition, systems with
multiple species and/or spinor bosons have also been studied; the
low-energy physics of their Mott phases are mostly controlled by
effective models arising out of an order-by-disorder mechanism. The
effective Hamiltonians obtained for such bosons may be described in
terms of interacting spins which may represents either real spin or
species degrees of freedom. These Hamiltonians lead to several spin
(or species) ordered bosonic ground states
\cite{girvin1,demler1,kush1,kush2,arun1,galit1}. However,
experimental realization of such strongly correlated states of
bosonic systems have not been yet achieved.

Instead, such a symmetry broken Mott phase was experimentally
realized through a slightly unexpected route. The key idea was to
generate an "electric field" for these neutral bosons \cite{bref1,
bref2, bref3}. As shall be detailed later, such a field may be
generated experimentally by shifting the center of the trap which
confines the bosons leading to a linear potential term in their
Hamiltonian. Alternatively, it can also be realized by application
of a linearly spatially varying Zeeman magnetic field to spinor
bosons. The presence of such a field leads to the realization of
bosonic Mott ground states with broken $Z_2$ symmetry \cite{subir2,
bref2}. In addition, it has recently been realized that the model
used \cite{subir2} to describe such systems can also describe the
physics of Rydberg chains \cite{gr1,gr2,gr3,gr4}. The study of
non-equilibrium dynamics of these Rydberg atoms has shown that they
may host several anomalous features \cite{scarlitrev,
scarlit1,scarlit2,scarlit3,scarlit4,scarlit5,ryddyn1,ryddyn2,ryddyn3}
including scar-induced dynamics \cite{scarlitrev,
scarlit1,scarlit2,scarlit3,scarlit4,scarlit5,
scarothers1,scarothers2,scarothers3,scarothers4} and possibility of
drive induced tuning of ergodicity properties
\cite{ryddyn1,ryddyn2,ryddyn3}. The main aim of this chapter is to
provide a summary of some of the theoretical and experimental
aspects of this rapidly developing field.

The rest of this chapter is organized as follows. In Sec.\
\ref{exp}, we provide a brief discussion of the experimental
platforms that lead to the realization of such tilted bosons. This
is followed by Sec.\ \ref{model} where we describe a low-energy
model which describes the ground states phases of these system and
discuss its relation with models describing a chain of ultracold
Rydberg atoms. This is followed by Sec.\ \ref{dynamics} where we
discuss non-equilibrium dynamics of the model. Next in Sec.\
\ref{fut} we discuss possible related models with interesting
properties which may be realized using such boson platforms.
Finally, we conclude in Sec.\ \ref{conc}.

\section {Experimental Platforms}
\label{exp}

In this section, we shall describe the essential ingredients of
three experimental setups. The first two involves generating a tilt
for ultracold bosons in a optical lattice while the third involves
Rydberg atoms in one-dimensional (1D) lattice.

The first experiment on tilted bosons in an optical lattice was
carried out in Ref.\ \onlinecite{bref1}. In this experiment, spinor
$^{87}{\rm Rb}$ atoms in their angular momenta $F=2$ and $m_F=2$
state (where $F$ is the total and $m_F$ is the azimuthal angular
momenta) were cooled in a trap. The trap was chosen to be a cigar
shaped magnetic trap with radial and axial frequencies $\nu_{\rm
radial}= 240 \,{\rm Hz}$ and $\nu_{\rm axial}=24 \,{\rm Hz}$. The
trap included $2 \times 10^5$ bosonic atoms. After the condensate
was the formed, the radial trapping frequency was relaxed to $240 \,
{\rm Hz}$ over a time period of $500$ms. This led to a spherically
symmetric condensate.

To generate an optical lattice, six orthogonal lasers with
wavelength $\lambda=852$nm were applied. This resulted in a
potential of
\begin{eqnarray}
V(x,y,z) &=& V_0 (\sin^2 k x + \sin^2 k y + \sin^2 k z )
\label{oppot}
\end{eqnarray}
where $k= 2 \pi /\lambda$ and $\lambda$ is the wavelength of light
used for the lasers. For such optical lattices, all energies are
typically measured in terms of the recoil energy given by $E_r=
\hbar^2 k^2/(2m)$; this is the basic energy scale in the problem
which can be created out of the wavelength $\lambda$ of the laser
and mass $m$ of the atoms. It can be shown that in such a lattice,
the atoms see an approximately harmonic potential with strength $V'
\simeq \sqrt{V_0 E_r}$ leading to a trapping frequency $\nu_r \simeq
V'/h \simeq 30$kHz. In the experimental setup of Ref.\
\onlinecite{bref1}, the strength of the trapping potential could be
up to $V'= 22 E_r$. These values of parameters were sufficient to
obtain a Mott insulating state of bosons with one boson per site of
the optical lattice.

In addition, to apply the tilt, the center of the trap confining the
bosons was shifted. This led to a shifted harmonic potential. The
bosons thus sees a linear gradient since $ V_{\rm shifted}= K_0 (x-
x_0)^2/2 \simeq V_{x_0=0} - c x $, where $c=K_0 x_0$, $x_0$ is the
shift of the trap center, and we have ignored an irrelevant constant
term. Thus the shift is analogous to having an electric field for
the neutral bosons with $eE = K_0 x_0$; the magnitude of the field
can be controlled by controlling the shift. It was found in Ref.\
\onlinecite{bref1} that in the Mott phase, the presence of such a
shift leads to resonant energy absorption at special values of
electric field $eE$ (or shift $x_0$) which satisfies $eE = n U$
where $U$ is the on-site interaction between the bosons and $n$ is
an integer.

The measurement which confirmed this resonant absorption in the
experiments of Ref.\ \onlinecite{bref1} involved several steps.
First, the condensate was subjected to optical lattice potential
whose amplitude was slowly ramped up (over a period of $80$ms) to
the final value. This value is chosen ($V'= 22 E_r$) such that the
system would be in its Mott state with one boson per site. Second,
the system is allowed to equilibrate in this potential for a period
of $20$ms. During this time, the tilt of a fixed magnitude is
applied to the system. Third the optical lattice potential is
reduced to $V' = 9 E_r$ (for which the bosons are in a superfluid
state) within a short time interval of $3$ms. Finally, the trap and
lattice is turned off, and the real space imaging of the flying out
bosons are carried out. It is to be noted that after turning off the
trap and the lattice, the bosons undergo a free flight. Thus their
position distribution during the flight provides information about
their momentum distribution (or initial velocity distribution)
inside the trap just before it was turned off. Consequently, in the
superfluid phase, the position distribution of the bosons would have
a large central peak signifying the presence of a large number of
bosons in the state $\vec k=0$ inside the trap. The experiment in
Ref.\ \onlinecite{bref1} further noted that if the system absorbed
energy when the tilt is applied, it is not going to equilibrate to
the superfluid ground state when the lattice potential is reduced.
Instead, it will be in an excited state where the boson wavefunction
have larger weight in the $\vec k \ne 0$ states in the trap. This
will broaden their position distribution leading to a broader
central peak during imaging. A large width of the central peak of
the image therefore is a signature of large energy absorption due to
the perturbation; such a large width is observed around $eE = nU$.

The measurement technique of Ref.\ \onlinecite{bref1} did not allow
for a direct measurement of the number distribution of the bosons
within the lattice in the Mott phase. This is clearly desirable if
one wanted to distinguish between several competing ground states in
the Mott regime. The later experiments, performed in Ref.\
\onlinecite{bref2, bref3}, made significant progress in this
direction. In these experiments, which also used $^{87}{\rm Rb}$
atoms, lasers with wavelength of $\lambda= 680$nm were used to
generate the optical lattice. The trap used to confine the bosons
was also optical. The maximum allowed lattice depth achieved in
these experiments were $V'= 45 E_r$ which brought the bosons close
to their Mott state in the atomic limit. The lattice obtained was a
2D lattice; however, the experiment had separate control over the
lattice depth in the $x$ (along the chain) and the $y$ (between the
chains) direction. The inter-chain lattice depth could be ramped to
very high to achieve an effectively 1D optical lattice (almost
disconnected chains).

The tilt generated in these experiments were carried out via
application of Zeeman magnetic field which varied linearly in space.
Such a Zeeman field leads to a potential term $ H_1 = - g \mu _B B_0
\sum _j j \hat n_j$ where $\mu_B$ is the Bohr magneton, $g$ is the
gyromagnetic ratio, $B_0$ is the amplitude of the field on the first
site ($j=1$) and $\hat n_j$ is the number operator for the bosons.
Such a field therefore creates an effective electric field for the
bosons  with $e E= g \mu_B B_0$. Note that the intensity of this
field can be controlled by tuning the magnetic field which is
experimentally much more convenient than shifting the trap center.

To measure the density distribution of the bosons inside a trap, the
experiments in Ref.\ \onlinecite{bref2,bref3} used an ingenious
fluorescence imaging technique. In this technique, the depth of the
lattice potential is suddenly increased just before the measurement
so that the bosons within the lattice freeze for a long time scale.
Then fluorescent light is applied to the bosons; the frequency of
this light is chosen in such a way that any boson pair on a lattice
site can scatter via light assisted collision and move out of the
trap. Thus such a fluorescent light leaves behind an empty lattice
site if there were, initially, an even number of bosons on that
site; in contrast, if there are odd number of bosons, at least one
boson remains on the site. Thus an imaging of the bosons after the
fluorescent light is applied provides information about their parity
of occupation in the Mott state. If a boson remains on a lattice
site, it may scatter several photons leading to bright spots. Thus
it provides a direct distinction between Mott states with even and
odd occupations (for example between $0-2-0-2...$ and $1-1-1-1...$
occupations) of bosons; the sites with even occupations lead to dark
spots while that with odd occupation appear bright. This proves to
be very useful while measuring boson occupation in the presence of a
tilt.

Finally we briefly discuss some experiments on ultracold bosonic
Rydberg atoms in a 1D optical lattice \cite{gr1,gr2,gr3,gr4}. The
atoms used for such experiments are again $^{87} {\rm Rb}$. These
atoms are confined in a 1D optical lattice as discussed earlier. In
these experiments, the atoms are subjected to a Raman laser which
induces a transition between the ground ($|g\rangle = |5S_{1/2};
F=2; m_F=-2\rangle$) and the Rydberg excited ($|r\rangle = |70
S_{1/2}; F=1/2; m_F=-1/2\rangle$) state of the atoms via an
intermediate state ($|p\rangle = |6P_{3/2}; F=3, m_F=3\rangle$). The
experiment used two lasers with wavelength $420$nm and $1024$nm
(corresponding to single photon Rabi frequencies of $\Omega_B= 2 \pi
\times 36$ and $\Omega_R= 2 \pi \times 60$ MHz respectively) so that
there is a detuning $\delta$ between the $|g\rangle$ and the
$|p\rangle$ levels: $\hbar \Omega_B= \delta +(E_p-E_g)$. Similarly
the detuning $\Delta$ between the $|r\rangle$ and the $|g\rangle$
levels are given by $\Delta= \hbar(\Omega_B +\Omega_R) - (E_r-E_g)$.
For $\Delta=0$ the atoms would be in a equal linear superposition of
$|g\rangle$ and $|r\rangle$ states. $\Delta$ can be tuned in
experiments to have either positive or negative values. The
effective coupling between the atoms in the ground and Rydberg state
is given by $\Omega= \hbar^2 \Omega_B \Omega_R/ (2 \delta)$. Further
experimental details regarding the setup can be found in Ref.\
\onlinecite{gr1}.

The atoms experience a strong repulsive dipolar interactions ($V(r)
\sim 1/|r|^6$) when excited in the Rydberg state. This interaction
can be tuned by controlling the relative position of the atoms in
the lattice; in particular, a regime can be reached which precludes
two Rydberg excited atoms within a certain length $R$. This
phenomenon is called the dipole blockade and $R$ is termed as the
blockade radius. In experiments, it is possible to control the
dipolar interaction strength between the atoms leading to an
effective tuning of the blockade radius. Such a tuning of the
blockade radius may lead to translational symmetry broken phases as
follows. In experiments the blockade radius can be tuned to next
neighbor; moreover, the parameters are so adjusted that it is
energetically favorable for each individual atom to be in its
Rydberg excited state. A competition between these two phenomenon
leads to a state where atoms in every alternate site can be in the
state $|r\rangle$ leading to a symmetry broken state. Similarly
states with blockade radius of two lattice sites can be achieved;
these states, for large negative $\Delta$, shall have a Rydberg
excited in every three sites. In the next section, we shall find
that the physics of such a system can be understood in terms of a
model which has several common features with the tilted Bose Hubbard
model.

\section{Model and Phases}
\label{model}

The low-energy effective model for the bosons in a tilted 1D optical
lattice has been derived in Ref.\ \onlinecite{subir2}. An extension
of this model has been studied in Ref.\ \onlinecite{fendley1}. In
the first part of this section, we briefly sketch the method of
derivation of this model from the microscopics. In the next part, we
shall discuss the similarity of the model with a spin model
appropriate for describing the Rydberg atoms discussed in the
previous section. The extension of this model to higher dimension
\cite{subir4,others1,bhaskar1a}, its application to tilted dipolar
bosons \cite{others2,others3}, tilted spin chains \cite{others4},
and other models with modified constraint \cite{bhaskar1b} has also
been worked out; however, in the rest of this work, we shall
restrict ourselves to the initial 1D model proposed in Ref.\
\onlinecite{subir2}.

\begin{figure}[ht]
\centering
\includegraphics[width=\linewidth]{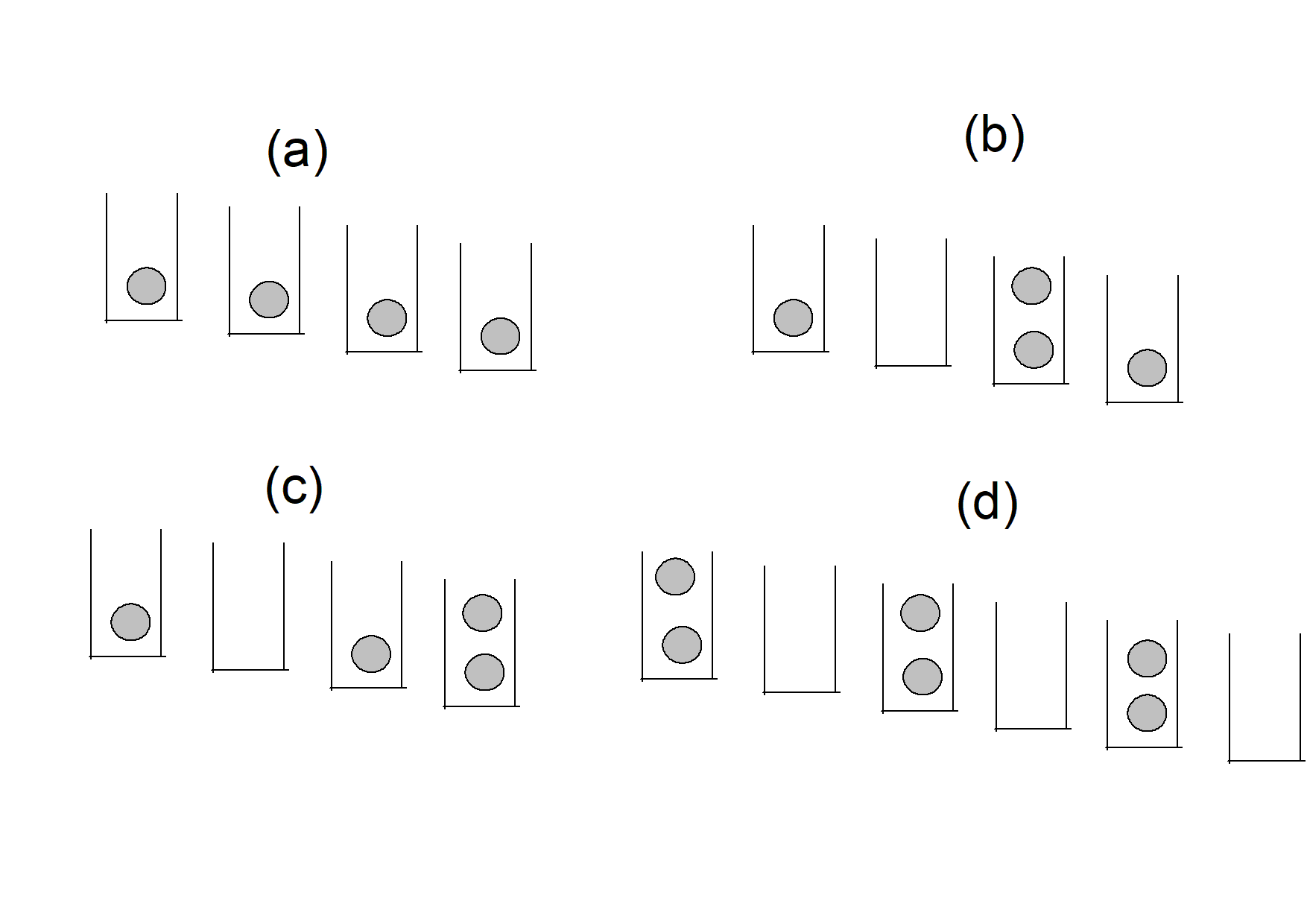}
\caption{(a) Schematic representation of the parent Mott state with
$n_0=1$. (b) The state with a single dipole. (c) A state with two
neighboring dipoles which is equivalent to a single dipole of length
two and is not a part of the low energy subspace. (d) A state with
maximal number of dipoles which is also a part of the low-energy
subspace.} \label{fig1}
\end{figure}

In the presence of a tilt, the Hamiltonian of ultracold bosons in a
1D optical lattice is given by
\begin{eqnarray}
H_b &=& - w \sum_{\langle j j'\rangle} (b_j^{\dagger} b_{j'} + {\rm
h.c.}) - \mu \sum_j \hat n_j +\frac{U}{2} \sum_j \hat n_j(\hat
n_j-1) - {\mathcal E} \sum_j j \hat n_j  \label{micham}
\end{eqnarray}
where $b_j$ denotes the boson annihilation operator at site $j$,
$\hat n_j= b_j^{\dagger} b_j$ is the boson number operator, $w$ is
their nearest-neighbor hopping amplitude, $U$ is the amplitude of
the on-site interaction between the bosons, and $\mu$ denotes their
chemical potential. In Eq.\ \ref{micham}, $\langle j j'\rangle$
indicates that $j$ and $j'$ are nearest neighbors, and ${\mathcal
E}$ denotes amplitude of the applied electric field in units of
energy. In absence of the field, the bosons are in the in Mott state
so that $\mu, U \gg w$; moreover in what follows, we shall address
the physics of the system for $U, {\mathcal E} \gg w, |U- {\mathcal
E}|$.

Before embarking on the description of the ground state of this
model, it is useful to think about the limit $U=0$. In this case the
model represents Wannier-Stark ladder for the bosons. The
single-particle Schrodinger equation for such non-interacting bosons
is given by (with energies $\epsilon= E+ \mu$)
\begin{eqnarray}
\epsilon \psi_j &=& -{\mathcal E} j \psi_j -w \left[\psi_{j-1}+
\psi(j+1) \right] \label{wseq}
\end{eqnarray}
The solution to this problem is well-known \cite{wsref}. The energy
of the bosons are given by $\epsilon_m = {\mathcal E} m $ where $m$
denotes integers ranging from $-\infty$ to $\infty$. Note that the
energy is not bounded from below and this arises from the fact that
the potential is also not bounded on an infinite lattice. The
eigenfunctions of the bosons are given by
\begin{eqnarray}
\psi_m (j,t) = J_{j-m}(2w/{\mathcal E}) \exp[-i {\mathcal E} m
t/\hbar], \label{starkeq}
\end{eqnarray}
where $J_n(x)$ denotes the Bessel function. Note that the
wavefunction returns to its original value at regular time intervals
$t_0= 2 \pi \hbar/\mathcal E$ which indicates Bloch oscillations.
The bosons are strongly localized to their respective site for $w\ll
{\mathcal E}$. This behavior can also be understood from the fact
that $J_{j-m}(x)$ has appreciable weight at $j=m$ for $x\ll 1$
leading to more than exponentially localized wavefunctions
\cite{subir2}. This behavior is in contrast to the classical
expectation where an electric field shall accelerate the bosons to
the end of the chain causing an electric breakdown. This feature
shall be a key component in constructing the effective theory in the
limit ${\mathcal E} \gg w$. Such a breakdown can happen in realistic
quantum systems due to electric field assisted tunneling to higher
single particle bands; however for ultracold bosons, such bands are
well separated in energy from the lowest bands. This leads to a
tunneling time which is larger than the typical system lifetime and
allows us to ignore such electric breakdown. In the absence of such
breakdown, the system remains in the metastable parent state
$|\psi(t=0)\rangle$. Thus the strategy adapted in Ref.\
\onlinecite{subir2} for finding out the equilibrium phase of the
tilted bosons was to start from the metastable parent Mott state and
find out the manifold of states whose energies are close to the
parent Mott state in the regime ${\mathcal E}, U \gg w, |{\mathcal
E}-U|$.

To understand the nature of the effective theory of the interacting
bosons, we therefore start from the parent Mott state (the Mott
ground state in the absence of the electric field) with $n_0$ bosons
on every lattice site (see Fig.\ \ref{fig1}(a)) and ask how the
electric field may destabilize such a bosonic system. A possibility
is that this can be achieved by addition of an extra particle or
hole in over the parent Mott state. The energy cost of adding a
particle to the Mott state is $E_p = U (n_0+1)-\mu$ while that for a
hole is $E_h = \mu-U n_0$. Note that $E_p, E_h >0$ as long as the
system is in the Mott ground state. However, once added, the
particle[hole] sees the electric field and is thus described by an
effective Hamiltonian which is identical to Eq.\ \ref{wseq} with
$\mu=0$ and $ w \to w(n_0+1)[w n_0]$. They are therefore localized
and can not reduce their energy via hopping (which would have been
the case if ${\mathcal E}$ was absent). Thus such excitations are
not effective in destabilizing the Mott states. This situation is
therefore different from that of bosons without the tilt, where such
additional particles or holes destabilize the Mott state to bring
about the superfluid-insulator transition \cite{free1, nicolas1}.

The excited states which are part of the low-energy manifold around
the parent Mott state correspond to dipole excitations where a
particle hops from a lattice site to the next (Fig.\ \ref{fig1}(b))
\cite{subir2}. Such an excitation creates an additional particle on
the site $j+1$ and a hole at site $j$ and thus have an energy cost
\begin{eqnarray}
E_{\rm dipole} &=& E(n_0+1)+E(n_0-1)-2E(n_0) -{\mathcal E} =
U-{\mathcal E} \label{dipen}
\end{eqnarray}
A state with one or multiple dipoles thus become a part of the low
energy manifold for $U \simeq {\mathcal E}$. These states are
schematically represented in Fig.\ \ref{fig1}(d). These dipoles live
on the link $\ell$ between two adjacent lattice site $j$ and $j+1$.
Identifying the parent Mott state as the dipole vacuum, the
operators describing the creation of such dipoles can be written in
terms of the boson operators as
\begin{eqnarray}
d_{\ell}^{\dagger}|{\rm vac}\rangle  &=& \frac{1}{\sqrt{n_0(n_0+1)}}
b_j b_{j+1}^{\dagger} |{\rm Mott}\rangle
\end{eqnarray}
We note the following properties of the dipoles. First, there can be
at most one dipole on any given link. This is seen by noting that
creating another dipole on the same link cost an energy $E_2 =
4U-2{\mathcal E}$ (for $n_0>1$). Thus a state with two dipoles on
the same link is not a part of the low energy manifold. Second, two
dipoles on adjacent links cost an energy $U-2 {\mathcal E}$ (Fig.\
\ref{fig1}(c)) (this is equivalent to a length two dipoles where a
boson hops two sites as shown in Fig.\ \ref{fig1}(c)) and thus such
states are also not a part of this manifold. In fact it can be shown
that all longer length dipoles do not play a role \cite{subir2}.
This leads to an effective dipole Hamiltonian supplemented by the
constraints
\begin{eqnarray}
H_d &=& -w' \sum_{\ell} (d_{\ell} + d_{\ell}^{\dagger}) + \lambda
\sum_{\ell} \hat n_{\ell}, \quad \hat n_{\ell} \le 1, \quad \hat
n_{\ell} \hat n_{\ell+1}=0 \ \label{diham}
\end{eqnarray}
where $\hat n_{\ell}= d_{\ell}^{\dagger} d_{\ell}$,
$\lambda=U-{\mathcal E}$ and $w'=w\sqrt{n_0(n_0+1)}$. We note that
the model does not conserve dipole numbers since spontaneous hopping
of bosons leads to formation/annihilation of dipoles. Moreover, the
model is non-integrable due to the presence of the second
constraint. The number of states in the Hilbert space of this
Hamiltonian does not scale as $2^L$ (as that of hardcore bosons or
spins); it can be shown that they scale as $\varphi^L$ for large $L$
where $\varphi= (1+\sqrt{5})/2$ is the golden ratio. It was shown
that it is possible to write down a recursion relation for the
Hilbert space dimension, ${\mathcal N}_L$, of the constrained dipole
system with periodic boundary condition: ${\mathcal N}_L={\mathcal
N}_{L-1}+{\mathcal N}_{L-2}$. This relates ${\mathcal N}_L$ to
Fibonacci numbers $F_L$ for integer $L$: $N_L = F_L$ \cite{ryddyn1}.

It is easy to see that $H_d$ admits a representation in terms of
spin-half Pauli matrices due to the constraint $\hat n_{\ell}\le 1$.
Indeed the mapping
\begin{eqnarray}
\sigma_{\ell}^z &=& 2\hat n_{\ell}-1, \quad \sigma_{\ell}^x =
(d_{\ell}+d_{\ell}^{\dagger}), \quad \sigma_{\ell}^y =
i(d_{\ell}-d_{\ell}^{\dagger})  \label{spintran}
\end{eqnarray}
leads to the spin Hamiltonian \cite{pekker1,scarlit2} (up to an
irrelevant constant)
\begin{eqnarray}
H_s &=& \sum_{\ell} (-w' \tilde \sigma_{\ell}^x + \lambda
\sigma_{\ell}^z /2 ), \quad  \tilde \sigma_{\ell}^a = P_{\ell-1}
\sigma_{\ell}^a P_{\ell+1}, \quad P_{\ell} = (1-\sigma_{\ell}^z)/2
\label{spinham}
\end{eqnarray}
Here we have implemented the constraint $\hat n_{\ell} \hat
n_{\ell+1}=0$ by using a local projection operator $P_{\ell}$ and it
is understood that $H_{\rm spin}$ operates in the constrained
Hilbert space where one can not have two up spins on the neighboring
link. The projection operator $P_{\ell}$ ensures that this
constraint is obeyed by $H_s$. We note here that the model, at
$\lambda=0$, has been dubbed as the $PXP$ model
\cite{scarlitrev,scarlit1,scarlit2,scarlit3,scarlit4,scarlit5}.

Next, we note that addition of longer-range density-density
interaction to $H_d$, leads to the Hamiltonian \cite{fendley1}
\begin{eqnarray}
H'_d  &=& H_d + V \sum_{\ell} \hat n_{\ell} \hat n_{\ell+2}
\label{fendham1}
\end{eqnarray}
We shall not discuss the details of the phases or the dynamics of
this model here but refer the readers to Refs.\
\onlinecite{fendley1,subir2a,subir3,ghosh1} for details. The model
displays a rich phase diagram and supports non-Ising quantum phase
transition. Moreover, this also serves the low-energy effective
model for the Rydberg atoms discussed in the last section in certain
limit; we shall detail this point towards the end of this section.

The ground phase diagram of $H_d$ can be understood in a
straightforward manner. For $\lambda \gg w'$($U \gg {\mathcal E}+
w'$), the dipole excitations are energetically costly and the ground
state is the dipole vacuum $|{\rm vac}\rangle$ (the parent Mott
state $|{\rm Mott}\rangle$). In contrast, for $|\lambda| \gg 0$ with
$\lambda <0$ (${\mathcal E} \gg U+w'$), the ground state is clearly
a state with maximal number of dipoles. However, due to the
constraint, there are two such maximal dipole states. The first
consists of dipoles are formed on the even links of the 1D lattice
while the second where they are formed on the odd links. These
states, for a chain of length $2L$ whose links are labeled from $0$
to $2L-1$, are (see Fig.\ \ref{fig1}(d))
\begin{eqnarray}
|{\mathbb Z}_2 \rangle &=& d_{0}^{\dagger} d_{2}^{\dagger} ...
d_{2L-2}^{\dagger} |{\rm vac}\rangle, \quad  |{\bar {\mathbb
Z}_2}\rangle = d_{1}^{\dagger} d_{3}^{\dagger} ...
d_{2L-1}^{\dagger} |{\rm vac}\rangle, \label{fmstates}
\end{eqnarray}
The ground state chooses one of the two states and hence breaks
$Z_2$ symmetry. This implies that it must be separated from the
dipole vacuum ground state by a transition; this quantum phase
transition belongs to the Ising universality class and occurs at
\cite{subir2}
\begin{eqnarray}
{\mathcal E}_c = U+ 1.31 \sqrt{n_0(n_0+1)} w  \label{criteq}
\end{eqnarray}
Such a transition can be understood to be the result of competition
between the dipole number fluctuation arising from the first term in
Eq.\ \ref{diham} and the effect of the electric field in the second
term which makes such fluctuation energetically costly. A rough
estimate of ${\mathcal E}_c$ can also be obtained from a variational
wavefunction approach. The ordered state which breaks the $Z_2$
symmetry is characterized by an Ising order parameter. In the dipole
language this order parameter can be written as
\begin{eqnarray}
O &=& \frac{1}{2L} \sum_{\ell=0}^{2L-1} (-1)^{\ell} \hat n_{\ell}
\label{diord}
\end{eqnarray}
which is the dipole density at $k=\pi$ signifying the broken
translational symmetry of the ground state.

The presence of such a translational symmetry broken state was
directly verified in the experiment carried out in Ref.\
\onlinecite{bref2} using the fluorescence imaging technique
discussed earlier. The experiment started with a Mott state having
$n_0=1$ bosons per lattice site and generated a tilt using a
linearly varying Zeeman field. Thus the parent Mott state correspond
to an odd number of bosons per site and provided bright patterns in
an imaging measurement. In contrast, the maximal dipole state, which
constitutes an even number of states per site would provide a dark
image. It was found that an increase of the strength of the magnetic
field providing the tilt indeed led to a dark pattern; moreover, one
could coherently interpolate between such bright and dark imaging
patterns by tuning the strength of the magnetic field. This
constituted the realization of the symmetry broken Mott state for
ultracold bosons in an optical lattice.

Before ending this section, we briefly comment on the Rydberg atom
experiments and the relation of the Hamiltonian of such Rydberg
atoms with the model developed here. The low-energy effective
Hamiltonian of these atoms describes the physics of the system for
times which is smaller compared to typical decay scales of Rydberg
excited atoms. The Hamiltonian involves two states $|g\rangle$ and $
|r\rangle$. Defining $\tau_{\alpha}$ ($\alpha=x,y,z$) to be standard
Pauli matrices in the space of these two states, one can write
\begin{eqnarray}
H_{\rm Ryd} &=& -\sum_j (\Omega  \tau_j^x + \Delta \tau_j^z)  +
\sum_{j j'} V_{jj'} \hat n'_j \hat n'_{j'}  \label{rydham1}
\end{eqnarray}
where $\Delta$ and $\Omega$ has been defined in Sec.\ \ref{exp},
$V_{jj'}$ is the dipolar interaction between two Rydberg excited
atoms. Here $\hat n'_j= (1+\tau_j^z)/2$ denotes the number operator
for Rydberg excitations. We note that for $V_{j,j+1} \gg \Omega,
|\Delta|$ and $V_{j,j+n} \ll \Omega, |\Delta|$ for $n>2$, Eq.\
\ref{rydham1} reduces to Eq.\ \ref{fendham1} with $\Omega \to -w
\sqrt{n_0(n_0+1)}$, $\Delta \to \lambda$ and $V_{j,j+2} \to  V$.
Moreover if $V \ll \Omega, |\Delta|$, Eq.\ \ref{rydham1} reduces to
Eq.\ \ref{diham} (and hence to Eq.\ \ref{spinham}). It is therefore
expected that the ground state phase diagram of Rydberg atoms would
also reflect translational symmetry broken states for $\Delta <0$
and $|\Delta| \gg \Omega$.

In experiments carried out in Ref.\ \onlinecite{gr1}, $\Delta$ could
be tuned to large negative values to achieve the translational
symmetry broken states. Such states had a period $2$ when $V_{j,j+1}
\gg \Omega, |\Delta|$ and $V_{j,j+n} \ll \Omega, |\Delta|$ for
$n>1$, in accordance with the prediction of the dipole model. In
addition, it was possible to tune the strength of the Rydberg
interaction such that $V_{j,j+1}, V_{j,j+2} \gg \Omega, |\Delta|$
and $V_{j,j+n} \ll \Omega, |\Delta|$ for $n>2$. In this case, the
ground state broke $Z_3$ symmetry and led to a period $3$
density-wave in accordance with that found in Ref.\
\onlinecite{fendley1} by analyzing the Hamiltonian given in Eq.\
\ref{fendham1}.

\section{Non-equilibrium dynamics}
\label{dynamics}

In this section, we shall discuss the non-equilibrium dynamics of
the tilted bosons in the presence of an optical lattice. The quench
and the ramp dynamics of the bosons shall be discussed in Sec.\
\ref{qrdyn} while the periodic dynamics will be addressed in Sec.\
\ref{pdyn}.

\subsection{Quench and ramp protocols}
\label{qrdyn}

It is well-known that ultracold atoms provide a perfect platform for
studying non-equilibrium dynamics of closed quantum systems
\cite{rmpks,dziar1,adutta1,brefrev,bookchap1,bookchap2}. One of the
simplest protocol for such dynamics is the quench, where a parameter
in the Hamiltonian of the system is changed suddenly. Consequently,
the state of the system does not have time to react. The old ground
state of the system (or any initial state the system might be in
when the quench is performed) evolves according to the new
Hamiltonian. Since the old state is no longer an eigenstate of the
new Hamiltonian, it displays non-trivial dynamics. For
non-integrable systems, such dynamics is expected to be lead to a
thermal steady state at long times. This follows from the eigenstate
thermalization hypothesis (ETH)
\cite{ethref1,ethref2,ethref3,ethrev} which predicts eventual
thermalization of a typical many-body state under unitary dynamics.
At short times, the system is expected to display transient
oscillations.

\begin{figure}[ht]
\includegraphics[width=0.49\linewidth]{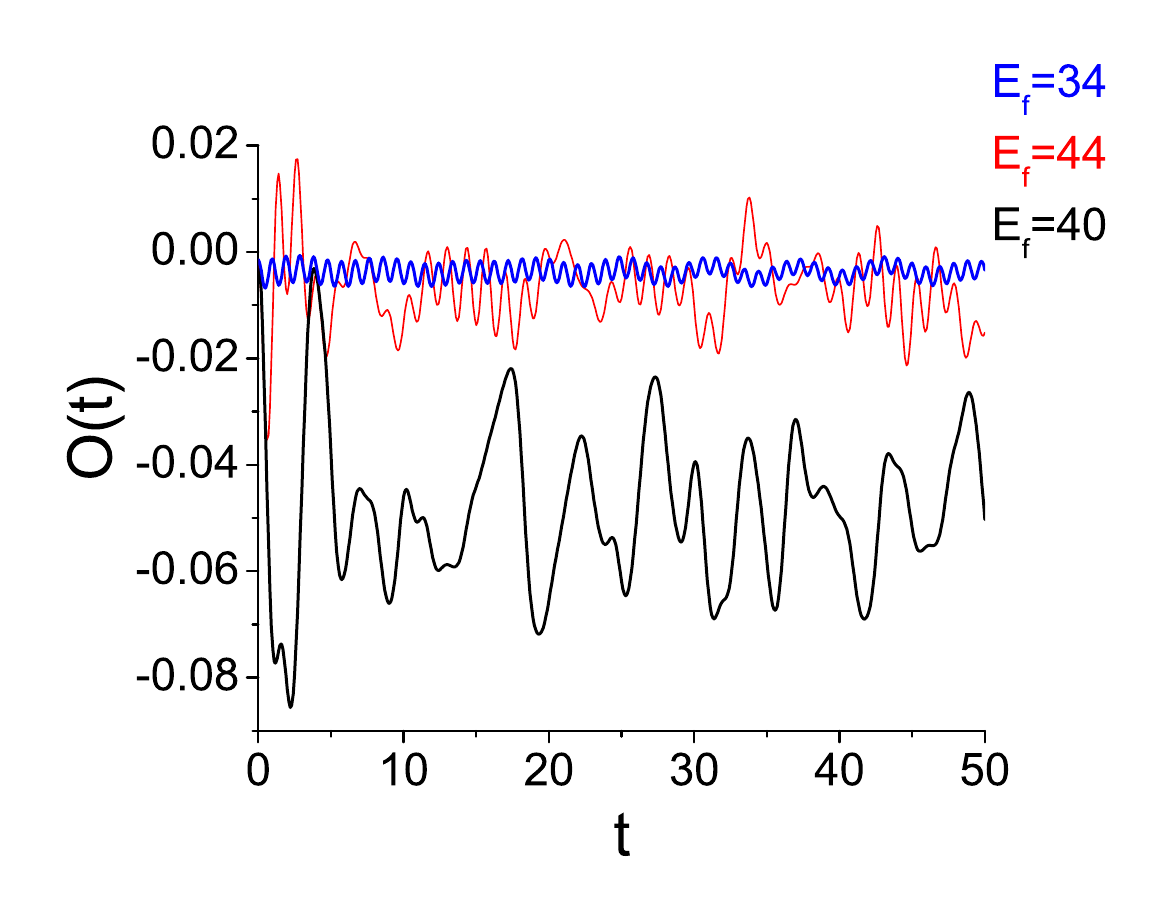}
\includegraphics[width=0.49\linewidth]{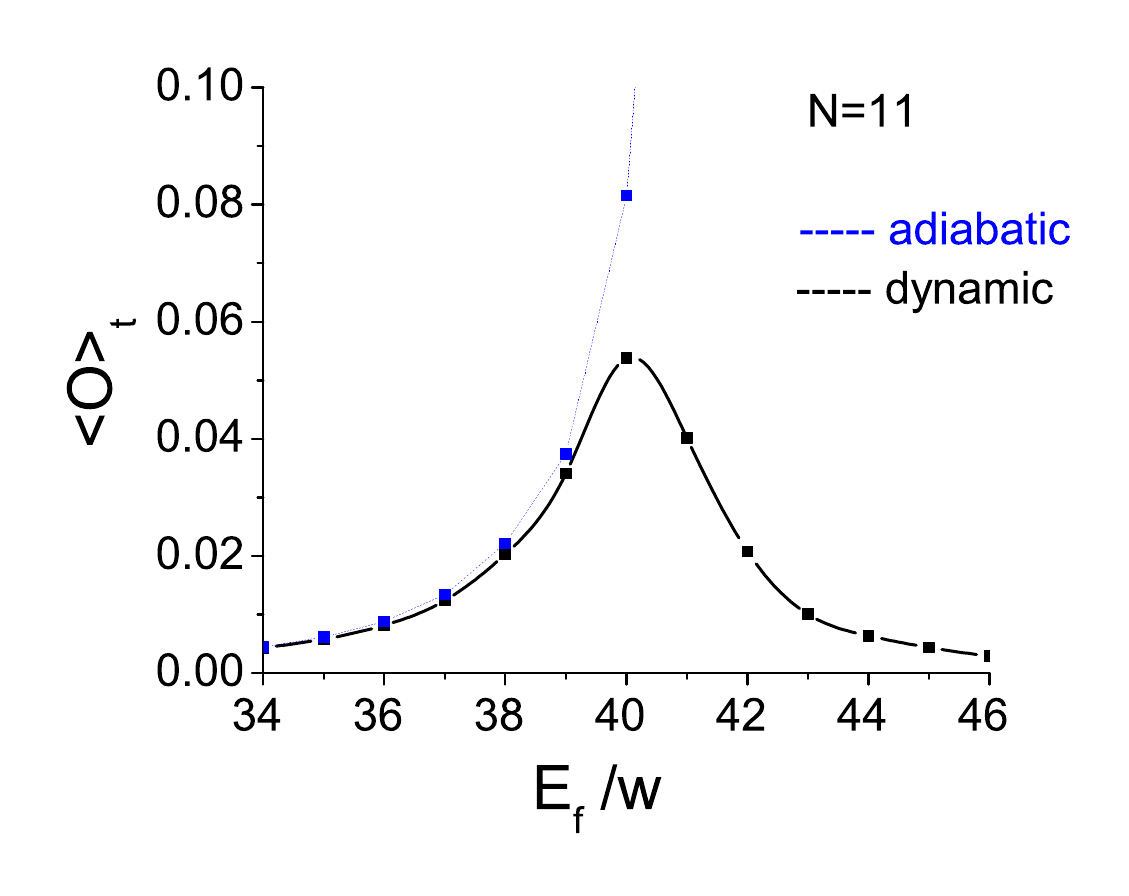}
\caption{Left panel: Plot of $O(t)$ as a function of time for
several representative values of ${\mathcal E}_f/w$. For all plots
$U/w=40$ and initial state is the dipole vacuum ground state
$|0\rangle$. Right panel: Plot of ${\bar O} \equiv \langle
O\rangle_t$ as a function of ${\mathcal E}_f/w$ showing a peak near
the critical point. In the figs $N$ represents the chain length in
units of lattice spacing. This figure is adapted from Ref.\
\onlinecite{ks1}.} \label{fig2}
\end{figure}

For the tilted boson system, such oscillations were studied in Ref.\
\onlinecite{ks1}. The bosons were initially assumed to a state of
dipole vacuum $|0\rangle$. At $t=0$, the electric field ${\mathcal
E}$ is changed from its initial value ${\mathcal E}_i < U$ to its
final value ${\mathcal E}_f$. The state of the system at time $t$
can then be described by
\begin{eqnarray}
|\psi(t)\rangle &=& \sum_n c_n e^{- i E_n t/\hbar} |n \rangle, \quad
H({\mathcal E}_f) |n\rangle = E_n |n\rangle, \quad c_n= \langle
n|0\rangle \label{stateqevol}
\end{eqnarray}
Under such evolution, the dipole order parameter given by Eq.\
\ref{diord} oscillates as
\begin{eqnarray}
O(t) &=& \sum_{m,n} c_m c_n O_{mn} \cos \omega_{mn} t , \quad O_{mn}
= \langle m|O|n\rangle, \quad \omega_{mn} = (E_m-E_n)/\hbar
\label{ordosc}
\end{eqnarray}
where we have used the fact that $c_m$ could be chosen to be real. A
numerical evaluation of $O(t)$, carried out in Ref.\
\onlinecite{ks1} (left panel of Fig.\ \ref{fig2}), showed that the
transient oscillations have maximal amplitude when ${\mathcal E}_f=
U$ near the critical point; they are tiny deep inside both the
maximal dipole and dipole vacuum phases as shown in the left panel
of Fig.\ \ref{fig2}. Thus transition can act as a qualitative marker
for change in state of the system.

To understand why the amplitude peaks around $U= {\mathcal E}_f$, we
focus on the time averaged value of these oscillations
\begin{eqnarray}
\bar O &=&  \lim_{T \to \infty} \frac{1}{T} \int_0^T O(t) = \sum_m
c_m^2 O_{mm}. \label{longtimreq}
\end{eqnarray}
The amplitude of these oscillations is maximal when the product of
overlap $c_m$ and order parameter expectation $O_{mm}$ is large.
When ${\mathcal E}_f$ is deep inside the dipole vacuum, $c_m \sim 1$
when $m$ correspond to the old ground state. However, $O_{mm} \to 0$
for this state leading to a small $\bar O$. In contrast, when
${\mathcal E}_f$ corresponds to deep inside the maximal dipole
ground state, $O_{mm}$ is large for the new ground state; however,
$c_m \sim 0$ for this ground state leading, once again, to a small
${\bar O}$. In between, near the critical point, ${\bar O}$ could be
large since both $c_m$ and $O_{mm}$ can be non-zero for several $m$.
This leads to a peak of ${\bar O}$ near the critical point as shown
in the right panel of Fig.\ \ref{fig2}.

More recently, the quench dynamics of Rydberg atoms has been studied
experimentally starting from the $|{\mathbb Z}_2\rangle$ state
\cite{gr1}. It was found that the evolution of such a state,
following a quench of the parameter $\Delta \to 0$, displays
long-lived coherent oscillations. Since the system is
non-integrable, ETH predicts that such dynamics will lead to an
eventual thermal steady state; however, such a steady state was not
observed in experiments for dynamics starting from $|{\mathbb
Z}_2\rangle$. In contrast, dynamics starting from the all spin-down
state ($|0\rangle$ or the dipole vacuum state) showed expected, ETH
predicted, thermalization. This phenomenon therefore constituted a
weak (initial-state dependent) violation of ETH in such finite-sized
chains.

The theoretical explanation of this phenomenon followed soon
\cite{scarlitrev,scarlit1,scarlit2,scarlit3,scarlit4,scarlit5}. The
details of this has been summarized in Ref.\ \onlinecite{scarlitrev}
and references therein. It was found at $\Delta=0$, or in the $PXP$
limit, the eigenspectrum of $H_s$ (or equivalently $H_{\rm Ryd}$ for
large $V_{i,i+1}$) supports a special class of eigenstates called
quantum scars
\cite{scarlitrev,scarlit1,scarlit2,scarlit3,scarlit4,scarlit5,
scarothers1,scarothers2,scarothers3,scarothers4}. These states have
finite energy density but anomalously low half-chain entanglement
entropy \cite{scarlitrev}. Being anomalous, they have very little
overlap with the thermal band of mid-spectrum eigenstates.
Consequently, they form an almost closed subspace. It was also found
that they have a strong overlap with the $|{\mathbb Z}_2\rangle$
state; thus dynamics starting from the $|{\mathbb Z}_2\rangle$ state
is almost confined within the closed subspace formed by the scars
leading to coherent long-lived oscillations. This does not happen
for dynamics starting from the $|0\rangle$ state since it has very
little overlap with the scar states. Further details of this
phenomenon and properties of quantum scars is summarized in Ref.\
\onlinecite{scarlitrev}.

\begin{figure}[ht]
\includegraphics[width=0.99\linewidth]{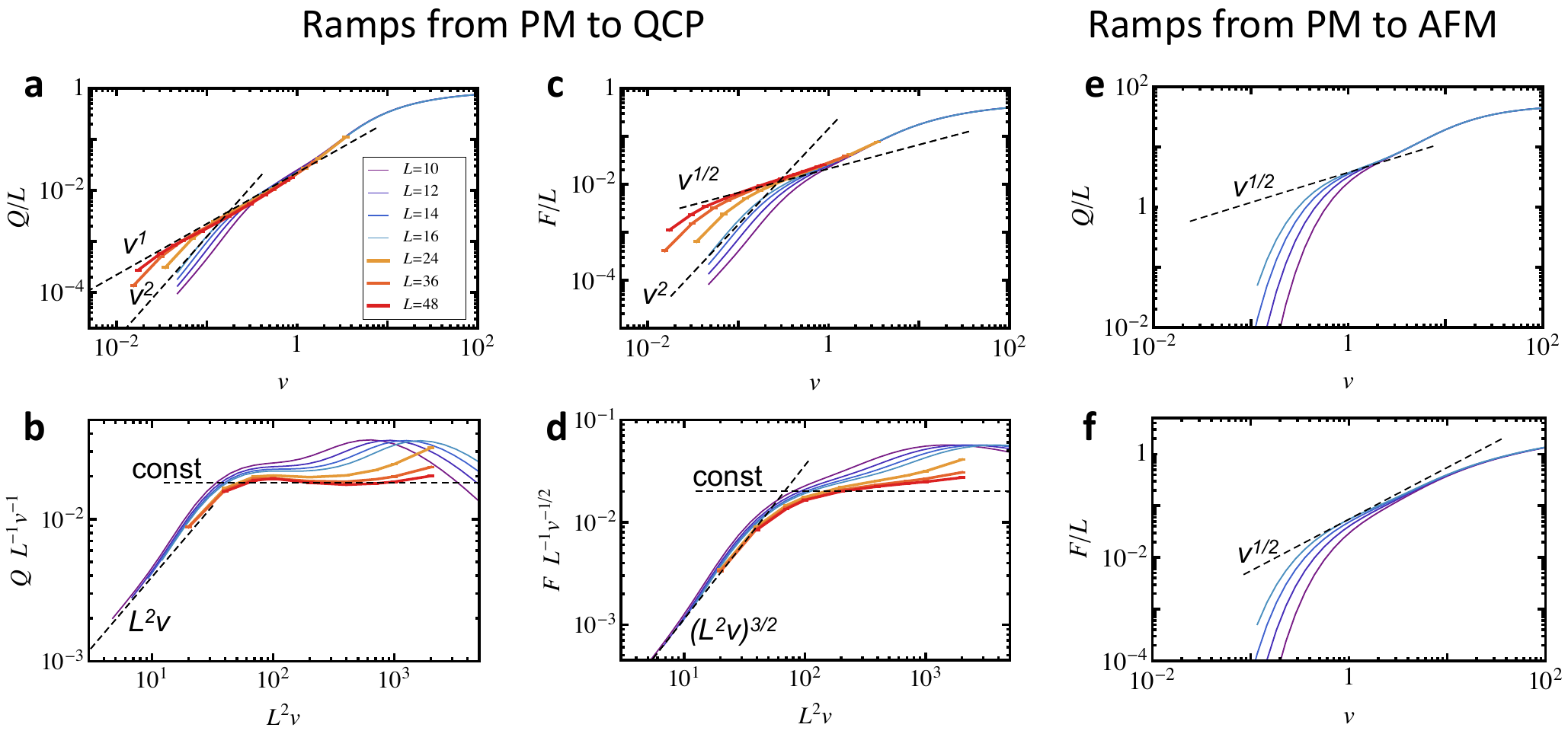}
\caption{Plots showing behavior of (a) residual energy $Q$ as a
function of $v=\tau^{-1}$ (b) $Q/(Lv)$ as a function of $v$ (c)
overlap $F$ as a function of $L^2 v$ and (d) $F/(L\sqrt{v})$ as a
function of $L^2 v$ for quench from the dipole vacuum or
paramagnetic ground state(PM) to the quantum critical point (QCP).
In the plot, all quantities are dimensionless, $U/w=40$, and the
lattice spacing $a$ has been set to unity. For ramp from the PM to
the maximal dipole or antiferromagnetic (AFM) state, the behavior of
$Q$ and $F$ are shown as a function of $v$ in panels (e) and (f)
respectively. This figure is adapted from Ref.\
\onlinecite{pekker1}.} \label{fig3}
\end{figure}

Next, we discuss the ramp dynamics of such a system addressed in
Ref.\ \onlinecite{pekker1}. In this case, one ramps the electric
field via a linear protocol from its initial value ${\mathcal E}_i$
at $t=0$ to a final value ${\mathcal E}_f$ at $t=\tau$ with a rate
$\tau^{-1}$: ${\mathcal E}(t) = {\mathcal E}_i +({\mathcal
E}_f-{\mathcal E}_i) t/\tau$. The values of ${\mathcal E}_f$ and
${\mathcal E}_i$ are so chosen so that the system starts from the
dipole vacuum state (ground state for ${\mathcal E}= {\mathcal E}_i
< U$) and reaches the critical point at ${\mathcal E}= {\mathcal
E}_f={\mathcal E}_c$. It is well known that such a passage to the
critical point leads to excitation production; the density of these
excitations, for slow ramp rates, scale with $\tau$ according to the
Kibble-Zurek (KZ) scaling law $n_{\rm ex} \sim \tau^{-\nu d/(z\nu
+1)}$, were $\nu$ and $z$ are the correlation length and dynamical
critical exponents and $d$ is the spatial dimension of the system
\cite{rmpks,dziar1,adutta1,bookchap1,bookchap2,kibble1,zureck1,anatoli1,ks2,ks3,anatoli2,sondhi1}.
However such scaling laws are strictly appropriate when the system
size is large. For finite-sized systems such as the Rydberg chain,
the system size $L$ provides a length scale which restrict the
applicability of the scaling laws.

\begin{figure}[ht]
\includegraphics[width=0.99\linewidth]{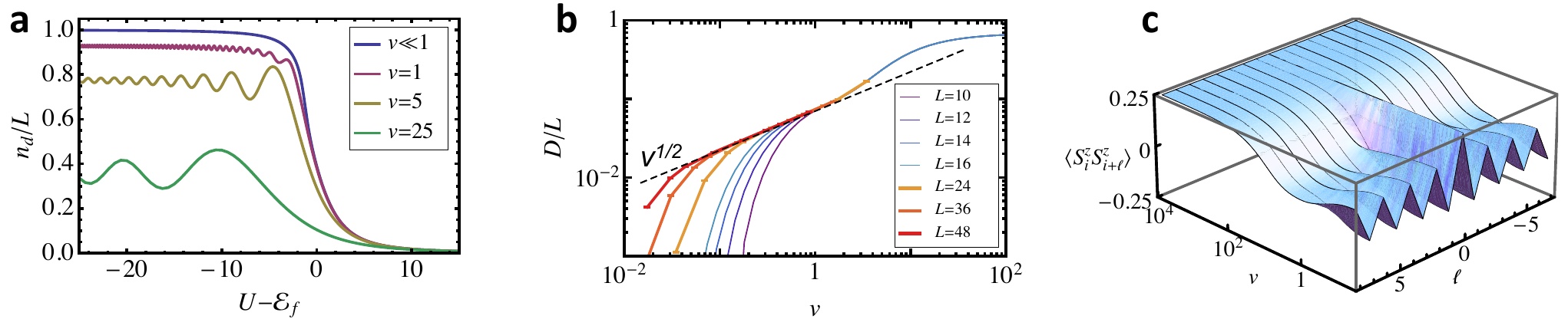}
\caption{Plots of (a) dipole density $n_d$ as a function of the ramp
amplitude $U-{\mathcal E}_f$ for several $v=\tau^{-1}$, (b)
excitation density $D$ as a function of $v$ when ${\mathcal
E}_f={\mathcal E}_c$, and (c) $C_{i i+\ell}$ as a function of $\ell$
and $v$. For all plots $U/w=40$ and the lattice spacing $a$ has been
set to unity. This figure is adapted from Ref.\
\onlinecite{pekker1}. } \label{fig4}
\end{figure}

To understand why the presence of a finite-system size changes
things, let us consider that the length scale $L$ leads to a energy
scale $\epsilon_0(L)$. When the ramp rate $\tau^{-1}  \le
\epsilon_0(L)/\hbar$, the system does not see the critical point;
instead the dynamics is similar to a two-level system (the two
states correspond to the instantaneous ground and first excited
states near the transition where excitations are formed) with
avoided level crossing, where the minimum gap is ${\rm
O}(\epsilon_0(L))$. In this case, the excitation density scales as
$n_{\rm ex} \sim \tau^{-2}$ and is independent of the critical
exponents; this is known as Landau-Zenner (LZ) scaling
\cite{anatoli3}. Also, when $\tau^{-1}$ is large, {\rm i.e.} for
fast drives, $n_{\rm ex}$ saturates and KZ scaling is not obeyed. In
between, there is a finite window of drive rates for which one finds
KZ scaling law. This behavior can be summed up by noting that for
finite-sized system, the scaling of excitation density is described
by
\begin{eqnarray}
n_{\rm ex} &=& N^{d} \tau^{-\nu d/(z\nu +1)} f(N^{1/\nu +z}
\tau^{-1}), \label{exden}
\end{eqnarray}
where $N=L/a$, $a$ is the lattice spacing, and the scaling function
$f(x)$ satisfies $f(x) = x^{2-\nu d/(z\nu +1)}$ for $x\ll 1$ and
$f(x)=1$ for $x\gg 1$. Note that $n_{\rm ex}$ crosses over from LZ
to KZ scaling regime with decreasing $\tau$ around $x \simeq 1$
$\tau \sim N^{2-\nu d/(z\nu +1)}$.

The dynamics of dipole chain for such ramps have been numerically
studied using exact diagonalization (ED) and time dependent matrix
product states (tMPS) methods. For ED, the analysis involves finding
the instantaneous eigenvalues and eigenvectors of the driven chain
at $t=t_f=\tau$; $H[t=\tau]|n\rangle = \epsilon_n |n\rangle$. One
can then expand the time dependent state $|\psi(t)\rangle = \sum_n
c_n(t) |n\rangle$, where $c_n(t)= \langle n|\psi(t)\rangle$. The
Schrodinger equation for the driven system can thus be written as
coupled differential equations for the coefficients $c_n(t)$ given
by
\begin{eqnarray}
i \frac{d c_n(t)}{d t} &=& \epsilon_n c_n(t) + \sum_{m}
\lambda_{mn}(t) c_m(t), \quad c_n(0) = \langle n|\psi(0)\rangle,
\quad \lambda_{mn}(t) = (E_f-E_i)(t/\tau-1) \langle m|\sum_j \hat
n_j|n\rangle  \label{rampeq1}
\end{eqnarray}
These equations need to be solved numerically to obtain
$|\psi(t)\rangle$. Having obtained $|\psi(t)\rangle$, one may
compute several quantities such as residual energy $Q$, log fidelity
$F$ (which is same as the excitation density), the dipole density
$n_d$ and the defect density $D$ given by
\begin{eqnarray}
Q &=& \langle \psi(\tau)|H(\tau)|\psi(\tau)\rangle -\epsilon_0,
\quad
F= \ln |\langle \psi(\tau)|0\rangle\rangle|, \quad
C_{\ell \ell'}= \langle \psi(\tau)| S^z_{\ell} S^{z}_{\ell'}|\psi(\tau)\rangle \nonumber\\
n_d &=& \langle \psi(\tau)|\sum_{\ell} \hat n_{\ell}
|\psi(\tau)\rangle, \quad  D = n_d - \langle 0|\sum_{\ell} \hat
n_{\ell}|0\rangle \label{def}
\end{eqnarray}
where $|0\rangle$ and $\epsilon_0$ denotes the wavefunction and
energy of the final ground state at $t=\tau$ and $S_{\ell}^z= 2 \hat
n_{\ell}-1$.

A plot of these quantities are shown in Fig.\ \ref{fig3} and
\ref{fig4} for several representative values of $L$. It was found
that they exhibit KZ scaling (with exponent $\nu d/(z\nu +1)= 1/2$
for $d=z=\nu=1$) within a finite window of ramp rate which depends
on $L$. As $L$ increases the KZ regime holds for slower ramp rates;
the crossover between the LZ and the KZ regime is shown by the sharp
drops in the figures. Thus these ramped boson chains can provide
experimental platform for testing KZ scaling.

Before ending this section, we would like to note that the ramp
dynamics for $H'_d$ (Eq.\ \ref{fendham1}) has also been studied in
details \cite{ghosh1,subir3}. Ref.\ \onlinecite{ghosh1} found KZ
scaling consistent with the presence of a $3$-state Potts transition
in these chains. However, work of Ref.\ \onlinecite{subir3} which
accessed larger chain lengths, has predicted the existence of
non-Ising like critical points with $z\ne 1$ (where $z$ is the
dynamical critical exponent) in such chains. Since $H'_d$ with large
$V$ is easily reproduced in Rydberg chains, these chains can also
act as platform for hosting such non-Ising quantum critical points.
We shall not discuss this issue further in this article but refer
the interested reader to Refs.\ \onlinecite{fendley1,subir3,ghosh1}.
Furthermore, the dynamics of tilted bosons in the presence of a
two-rate protocol \cite{sau1} has also been discussed; it was found
that such protocols may aid suppression of excitation formation in
these systems \cite{uma1}.

\subsection{Periodic protocols}
\label{pdyn}

In this section, we shall discuss the periodically driven tilted
boson chains \cite{ryddyn1,ryddyn2,ryddyn3}. For the rest of this
article, we shall explicitly use the spin representation and work
with $H_s$ (Eq.\ \ref{spinham}). The connection of the spins with
the original dipoles is given by Eq.\ \ref{spintran}. The methods
used shall be briefly discussed in Sec.\ \ref{permeth} while the
main results shall be presented in Sec.\ \ref{perres}.

\subsubsection{Methods}
\label{permeth}

The properties of a periodically driven system is best described in
terms of its Floquet Hamiltonian $H_F$ which is related to the
unitary time evolution operator $U$ by $U(T,0)= \exp[-i H_F
T/\hbar]$, where $T= 2 \pi/\omega_D$ is the time period of the
drive, $\omega_D$ is the drive frequency, and the wavefunction of
the driven system at any time $t$ is related to the initial
wavefunction by $|\psi(t)\rangle = U(t,0) |\psi(0)\rangle$. Thus the
stroboscopic dynamics of the system at $t=nT$, where $n$ is an
integer is completely controlled by its Floquet Hamiltonian
\cite{flref1,flref2,asenrev}

The Floquet Hamiltonian of any periodically can be computed by
comparing two equivalent expressions for $U(T,0)$
\begin{eqnarray}
U(T,0) &=& T_t \left \{ \exp\left[-i \int_0^T dt H(t) /\hbar\right]
\right\} = \exp \left[-i H_F T/\hbar \right] \label{fleq1}
\end{eqnarray}
For an interacting many-body system, it is usually not possible to
obtain $H_F$ exactly. This has led to several approximation schemes
for such computations \cite{asenrev}. In what follows, we shall
mostly use one of these schemes, namely, the floquet perturbation
theory (FPT) \cite{dsen1,tb1,ghosh2,asenrev}, to obtain the Floquet
Hamiltonian for the driven dipole chain. This method involves a
perturbation in drive amplitude; the term in $H(t)$ with the largest
amplitude is treated exactly, while the other terms are treated
using standard time-dependent perturbation theory \cite{asenrev}. In
contrast to the Magnus expansion technique, the drive period $T$
need not be a small parameter here; thus, the method allows us to
access the intermediate drive-frequency regime which can not be
accessed using Magnus expansion.

To study the properties of the periodically driven chain, we choose
to vary the electric field according to the square pulse protocol
such that
\begin{eqnarray}
\lambda(t) &=& -(+)\lambda_0/2 \quad {\rm for}\, \, t \le(>) T/2
\label{prot}
\end{eqnarray}
where $\lambda_0$ denotes the amplitude of the drive. In what
follows, we shall address the regime where $\lambda_0 \gg w$. In
this regime, the drive term is the one which has the largest
amplitude. Thus we write
\begin{eqnarray}
H_s(t) &=& H_{s0}(t) + H_{s1}, \quad H_{s0}(t) = \lambda(t) \sum_j
\sigma_j^z, \quad  H_{s1} = -w \sum_j \tilde \sigma_j^x
\label{driveham}
\end{eqnarray}
and treat $H_{s0}(t)$ exactly. Noting that $H_{s0}$ is diagonal in
the spin basis, we consider a complete set of states in the
constrained Hilbert space which has $m$ up-spins and denotes these
states as $|m\rangle$. Note that the positions of these spins (as
long as they are not nearest neighbors) do not change their
instantaneous energy under action of $H_{s0}$; thus each $|m\rangle$
represents a degenerate manifold of states. Using these states as
basis states one finds
\begin{eqnarray}
U_0(t,0) &=&  e^{i \lambda_0 t \sum_j \sigma_j^z /(2\hbar)} \quad
{\rm for} \quad t \le T/2
\nonumber\\
&=&  e^{i \lambda_0 (T-t) \sum_j \sigma_j^z/(2 \hbar)} \quad {\rm
for} \quad t \ge T/2
\end{eqnarray}
so that $\langle m| U_0 |n\rangle \sim \delta_{mn}$. We also note
that for $t=T$, $U_0(T,0)= I$ (where $I$ is the identity matrix);
thus $H_F^{(0)}=0$ for this protocol. This is a consequence of the
symmetric nature of the drive (Eq.\ \ref{prot}) which leads to a
vanishing average of $\lambda(t)$ over a drive cycle.

The first order perturbative correction to the evolution operator
$U$ can be obtained using standard time dependent perturbation
theory. This is given by
\begin{eqnarray}
U_1(T,0) &=& \frac{-i}{\hbar} \int_0^T dt U_0^{\dagger}(t,0) H_{s1}
U_{0}(t,0) \label{forder}
\end{eqnarray}
We note that $\tilde \sigma_j^x$ flips a spin on the $j^{\rm th}$
site, $\langle m| H_{s1}|n\rangle \sim \delta_{m,n\pm 1}$. Denoting
a state $|m +(-)\uparrow_j\rangle$ to be the one with one additional
(less) spin up residing at the $j^{\rm th}$ site, we can therefore
write \cite{ryddyn1,ryddyn2,ryddyn3}
\begin{eqnarray}
U_1(T,0) &=& \sum_m \sum_j \sum_{s_j =\pm 1} c^{(1)}_{s_j} |m\rangle
\langle m+s_j|, \quad c^{(1)}_s = \frac{4 i w}{\lambda_0} e^{i
\lambda_0 T s/(4\hbar)} \sin \lambda_0 T/(4\hbar)  \label{u1dyn}
\end{eqnarray}
where $s=\pm 1$. Noting that $\tilde \sigma^{\pm} |m\rangle = |m \pm
1$, and using Eq.\ \ref{fleq1} we get, to first order in
$w/\lambda_0$
\begin{eqnarray}
H_F^{(1)} &=& -w \frac{\sin \gamma}{\gamma} \sum_j (\cos \gamma
\tilde \sigma_j^x + \sin \gamma \tilde \sigma_j^y ), \quad \gamma=
\lambda_0 T/(4\hbar).  \label{forderfl}
\end{eqnarray}
We note that for $T\to 0$, $\gamma\to 0$, and in this limit
$H_F^{(1)}= -w \sum_j  \tilde \sigma_j^x = H_{\rm PXP}$. This is
also consistent with the Magnus result for $H_F$ which demands that
$H_F$ shall be average Hamiltonian given by
\begin{eqnarray}
H_{F}^{\rm magnus} &=& \frac{1}{T} \int_0^T H(t) dt = H_{\rm PXP}
\label{magnus}
\end{eqnarray}
Also, at this order, $H_{F}^{(1)}$ represents a PXP like Hamiltonian
up to a rotation and an overall renormalization by a factor $\sin
\gamma/\gamma$; it was noted in Ref.\ \onlinecite{ryddyn1} that this
result constitutes a resummation of a class of terms in the Magnus
expansion.

The higher order terms in the Floquet Hamiltonian can also be
computed. This has been carried out systematically in Ref.\
\onlinecite{ryddyn3} and leads to $H_F^{(2)}=0$ at second order.
This null result for $H_F$ to second order owes it existence to the
fact that there exists an operator $C = \prod_j \sigma_j^z$ which
satisfies $C U(T,0) C = U^{-1}(T,0)$ leading to $\{H_F , C \}=0$. As
shown in Refs.\ \onlinecite{scarlitrev,ryddyn1}, such
anti-commutation only allows for odd orders in the Floquet
Hamiltonian. At third order, a straightforward but tedious
calculation \cite{ryddyn3}
\begin{eqnarray}
H_F^{(3)} &=& -\alpha_0 \sum_j \left[\left( \tilde \sigma_{j+1}^+
\sigma_{j-1}^+ + \tilde \sigma_{j-1}^+ \tilde \sigma_{j+1}^+ \right)
\tilde \sigma_j^- - 6 \tilde \sigma_j^+ \right] + {\rm h.c.},
\nonumber\\
\alpha_0 &=& \left[e^{3i\lambda_0 T/(2\hbar)} + 3e^{i \lambda_0
T/(2\hbar)}(1 + i \lambda_0 T/\hbar ) + 2(1 - 3e^{i \lambda_0
T/\hbar} )\right] \frac{\hbar w^ 3 e^{-i \lambda_0 T/\hbar}}{3i
\lambda_0^3 T}. \label{tordfl}
\end{eqnarray}
The first term in $H_F^{(3)}$ involves three spin on neighboring
sites. We note that the form of this term, {\it i,e.}, the order in
which the $\tilde \sigma^{\pm}$ operators appear, is dictated by the
presence of the constraint; the order of appearance ensures that two
neighboring sites can never have two up-spins. The second term of
the Floquet Hamiltonian provides a shift to $H_F^{(1)}$ and
renormalize its coefficients. In what follows, we shall use this
perturbative Floquet Hamiltonian to understand the numerical
results.

The numerical approach to this problem, carried out in Ref.\
\onlinecite{ryddyn1,ryddyn3} used exact diagonalization(ED). The
first step in this direction involves numerical diagonalizaton of
$H[\pm \lambda_0]$. We denote the corresponding energy eigenvalues
and eigenvectors by
\begin{eqnarray}
H[\pm \lambda_0] |p_{\pm}\rangle &=& \epsilon_p^{\pm}
|p_{\pm}\rangle \label{ed1}
\end{eqnarray}
In terms of this, the evolution operator can be written as
\begin{eqnarray}
U(T,0) &=& e^{-i H[\lambda_0] T/(2\hbar)} e^{-i H[-\lambda_0]
T/(2\hbar)} = \sum_{p,q} c_{pq}^{-+} e^{-i
(\epsilon_p^-+\epsilon_q^+)T/(2 \hbar)} |p^-\rangle\langle q^+|,
\quad   c_{pq}^{-+} =\langle p^-|q^+\rangle \label{evolnum}
\end{eqnarray}
Thus one has a finite-dimensional matrix (for finite sized $L$)
which can then be diaonalized. Since $U$ is an unitary operator. its
eigenvalues are unimodular. Thus one can express it in term of its
eigenspectra as
\begin{eqnarray}
U(T,0) &=& \sum_{\alpha} e^{-i \epsilon_{\alpha}^F T/\hbar}
|\alpha\rangle \langle \alpha|, \quad  H_F |\alpha\rangle =
\epsilon_{\alpha}^F |\alpha\rangle \label{uop}
\end{eqnarray}
where $|\alpha\rangle$ are the eigenfunctions of $U(T,0)$,
$\epsilon_{\alpha}^F$ are the eigenvalues of the corresponding
Floquet Hamiltonian,  and the last equation follows from Eq.\
\ref{fleq1}. This procedure thus allows access to the exact Floquet
quasienergies and eigenfunctions for finite-sized systems. The
stroboscopic evolution, at $t=nT$ ( where $n$ is an integer), for
any operator ${\mathcal O}$, can thus be computed as \cite{ryddyn1}
\begin{eqnarray}
{\mathcal O}_n &=& \langle \psi(0)|(U^{\dagger}(T,0))^n {\mathcal O}
U^n (T,0) |\psi(0)\rangle = \sum_{\alpha, \beta} c_{\beta}^{\ast}
c_{\alpha} e^{-i(\epsilon_{\alpha}^F-\epsilon_{\beta}^F)T/\hbar}
\langle \beta|{\mathcal O}|\alpha\rangle \label{opexp}
\end{eqnarray}
where $|\psi(0)\rangle$ is the initial state and $c_{\alpha}=
\langle \psi(0)|\alpha\rangle$.

In what follows, we shall also be computing the half-chain
entanglement entropy of the Floquet eigenstates. The procedure for
this is as follows. First, corresponding to any eigenstate
$|\psi_n\rangle$, we construct a density matrix $\rho_n=
|\psi_n\rangle\langle \psi_n|$ which is defined on the full chain
with periodic boundary condition. Next we divide the chain into two
equal halves, $A$ and $B$,  with open boundary condition and trace
out the contribution of states residing in $B$. Thus each matrix
element of the reduced density matrix $\rho^A_n$ after tracing out
$B$ can be written as \cite{ryddyn1}
\begin{eqnarray}
\langle \rho_n^A \rangle_{\alpha \beta} &=& \sum_{\mu \in B
=1}^{{\mathcal N}_B} \langle \alpha; \mu| \rho_n |\beta; \mu \rangle
\label{redden}
\end{eqnarray}
where ${\mathcal N}_B$ denotes the Hilbert space dimension
corresponding to states residing in $B$ with open boundary condition
and the states $|\alpha\rangle$ and $|\beta\rangle$ have weights in
region $A$ \cite{ryddyn1}. While carrying out this procedure, one
has to be careful in excluding states where the right end of $A$ and
the left end of $B$ both has spin-up (or dipoles) since these states
were not part of the Hilbert space of the full chain owing to the
constraint. The half-chain entanglement $S_{L/2}^{(n)}$ can then be
obtained numerically using
\begin{eqnarray}
S^{(n)}_{L/2} &=& -\sum_{j=1}^{{\mathcal N}_A} q_{j} \ln q_{j}
\label{hce}
\end{eqnarray}
where $q_{j}$ denotes the eigenvalues of $\rho_n^A$ obtained via
numerical diagonalization. We shall use $S^{(n)}_{L/2}$ to
distinguish between states with volume- and area-law entanglement
entropies in the rest of this article.

\subsubsection{Results}
\label{perres}

To study the physics of the driven dipole chain, we concentrate on
half-chain entanglement $S_{L/2}$ (Eq.\ \ref{hce}) of the Floquet
eigenstates and the density-density correlator $\langle O_{j2}
\rangle = \langle \psi(nT)|\hat n_j \hat n_{j+2} |\psi(nT)\rangle$,
where $\hat n_j=(1+\sigma_j^z)/2$. In what follows, we shall discuss
the property of the stroboscopic evolution of $\langle O_{22}
\rangle$ ({\it i.e.} choosing $j=2$) starting from either the
$|{\mathbb Z}_2\rangle$ (antiferromagnetic state with up-spins on
even sites) or the $|0\rangle$ (ferromagnetic all spin-down state)
state. The evolution of $\langle O_{j j+2} \rangle$ for other values
of $j$ is identical as long as $j$ is chosen to be even
\cite{ryddyn1,ryddyn2,ryddyn3}

\begin{figure}[ht]
\includegraphics[width=0.49\linewidth]{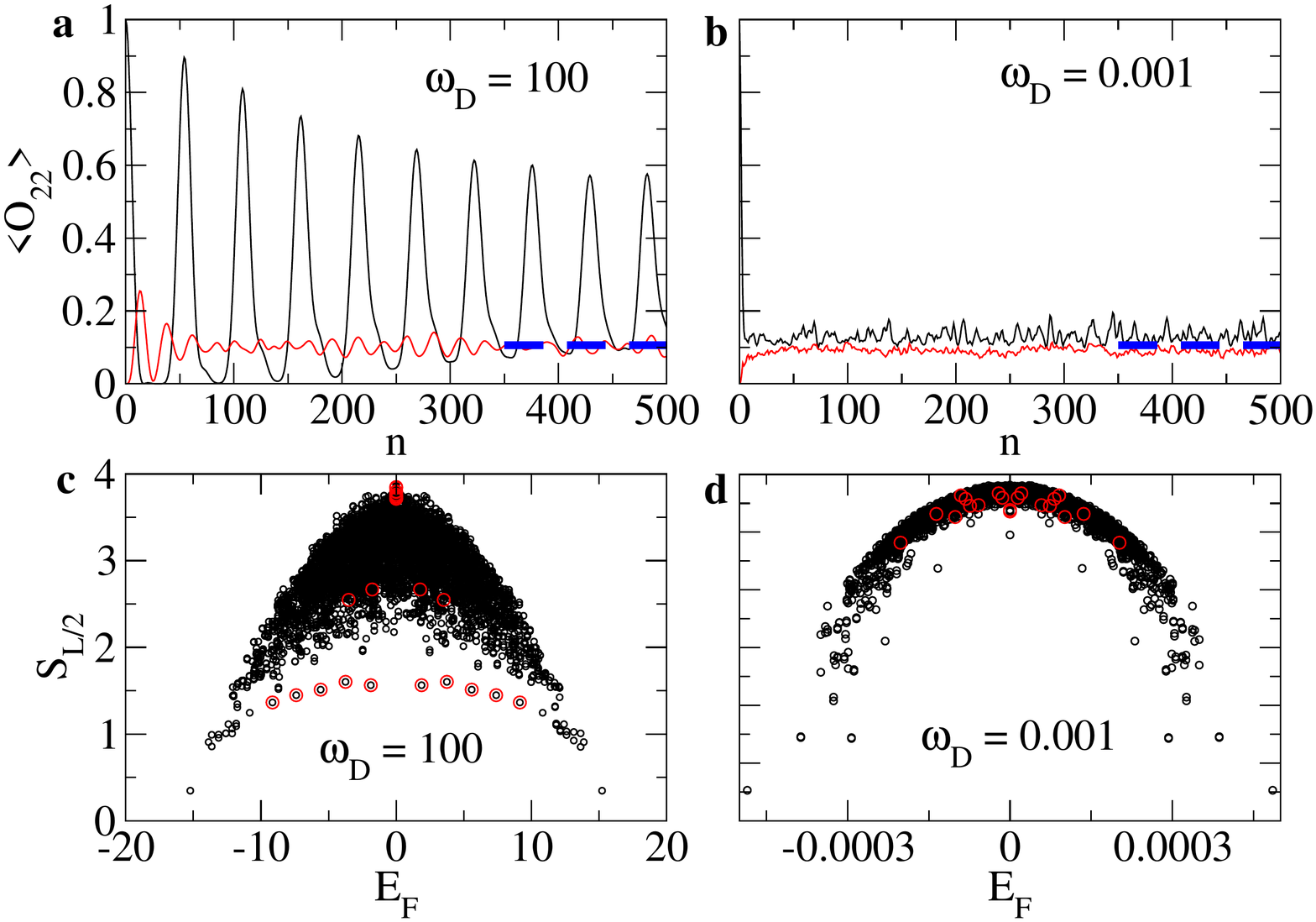}
\includegraphics[width=0.49\linewidth]{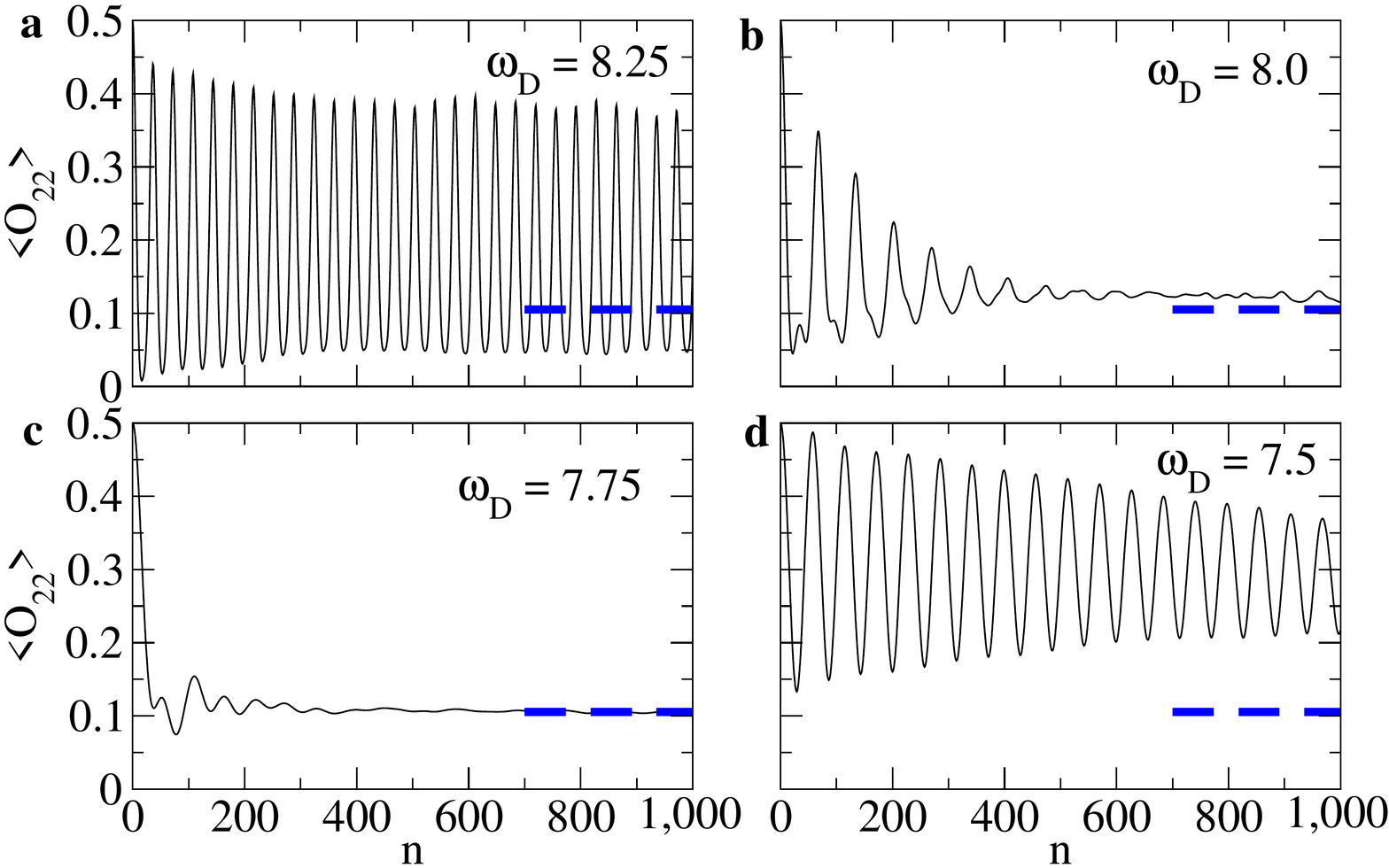}
\caption{Left Panel: Plot of $\langle O_{22} \rangle$ as a function
of number of drive cycles $n$ for (a) $\hbar \omega_D=100$ and (b)
$\hbar \omega_D=0.01$. The red lines indicates dynamics starting
from the state $|0\rangle$ while the black lines indicate $|{\mathbb
Z}_2\rangle$ The ETH predicted thermal steady state value of
$\langle O_{22} \rangle $ is $\sim 0.11$ as shown in blue dotted
line. Plots (c) and (d) shows the corresponding half-chain
entanglement $S_{L/2}$. Note the presence of athermal scar states
below the thermal band for $\hbar \omega_D=100$. Such states are
absent for $\hbar \omega_D=0.01$. The red circles correspond to
states which has high overlap with the initial $|{\mathbb
Z}_2\rangle$ state. For all plots $\lambda=15$, $w=\sqrt{2}$,
$\hbar$ is set to unity, and $L=18$. Right Panel: Plot of $O_{22}$
as a function of $n$ for (a) $\hbar \omega_D=8.25$, (b) $\hbar
\omega_D=8$, (c) $\hbar \omega_D= 7.75$, and (d) $\hbar \omega_D=
7.5$. The plot for (a) and (d) shows coherent oscillation while
those for (b) and (c) exhibits rapid approach to ETH predicted
thermalization. All other parameters are same that in the left
panel. This figure is adapted from Ref.\ \onlinecite{ryddyn1}.}
\label{fig5}
\end{figure}

We first consider the dynamics starting from the $|{\mathbb
Z}_2\rangle$ initial state. For this state, for $\lambda_0/(\hbar
\omega_D) \ll 1$, the dynamics is similar to that of quench studied
in Ref.\ \onlinecite{scarlitrev}. In this regime, we expect
scar-induced coherent oscillations with a frequency which is
determined by the energy separation between the quantum scar
eigenstates. In the opposite limit, $w/(\hbar \omega_D) \gg 1$
(where FPT and Magnus expansion both fail) it is expected that the
system shall heat up due to the drive and reach the thermal steady
state as predicted by ETH. This expectation is verified from the
behavior of $\langle O_{22}\rangle$ shown for high and low
frequencies in the plots shown at the top of the left panel in Fig.\
\ref{fig5}. The bottom plots of the left panel of Fig.\ \ref{fig5}
shows the half-chain entanglement entropies of the Floquet
eiegnstates at these drive frequencies; the high frequency
eigenstates clearly show the presence of athermal scars separated
from the thermal band of states. No such athermal states are seen at
low drive frequency; in this regime, all the states fall within the
thermal band. The red dots indicates eigenstates with large overlap
with the initial $|{\mathbb Z}_2\rangle$ state. These are the states
which primarily drive the dynamics. At high frequency, these states
are athermal and lie well outside the thermal band. Thus the
dynamics exhibit coherent oscillations since it involves a small,
${\rm O}(L)$, subspace of the full Hilbert space. In contrast, at
low frequency, the eigenstates having large overlap with $|{\mathbb
Z}_2\rangle$ are part of the thermal band leading to thermalization.

\begin{figure}[ht]
\includegraphics[width=0.49\linewidth]{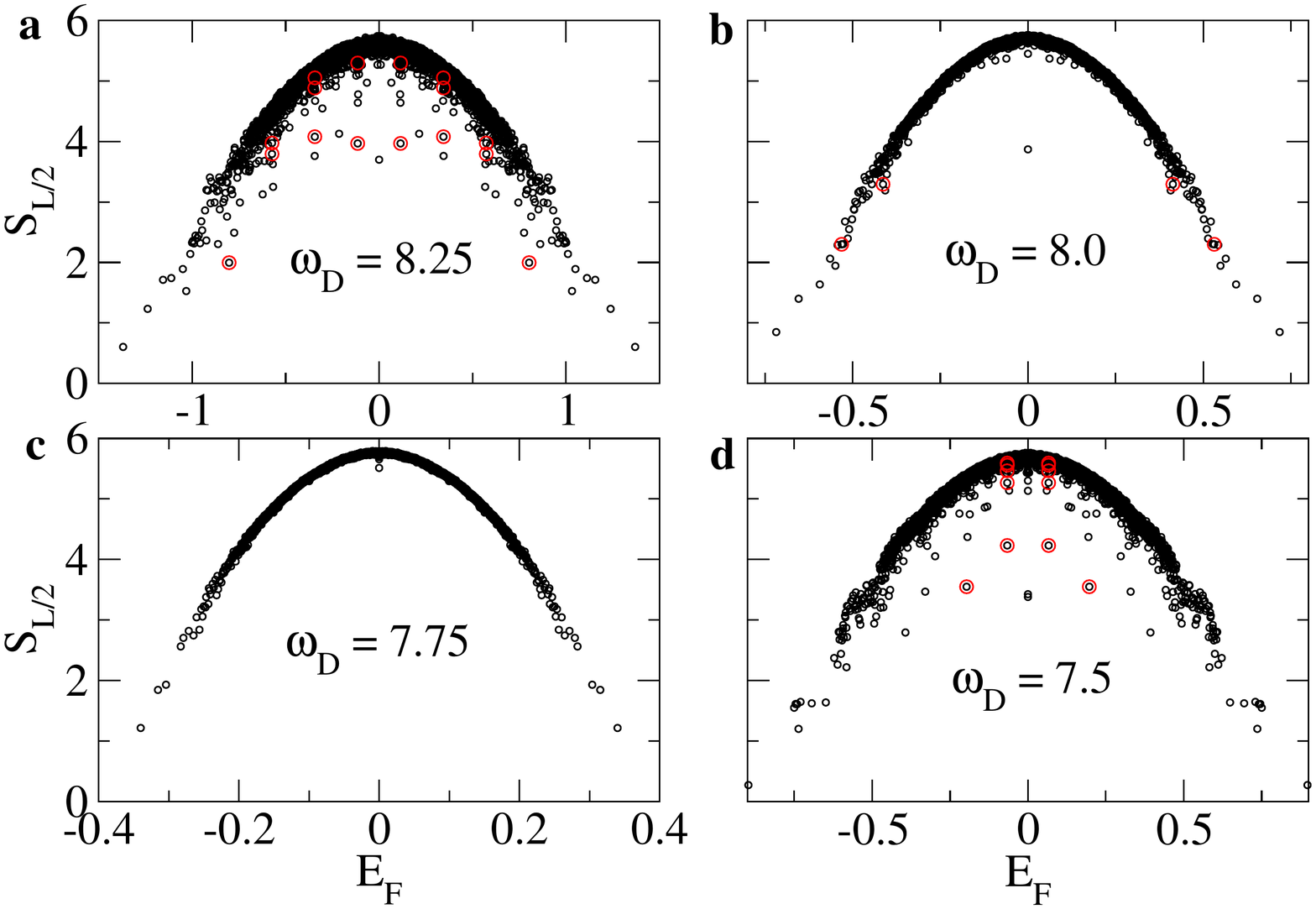}
\includegraphics[width=0.49\linewidth]{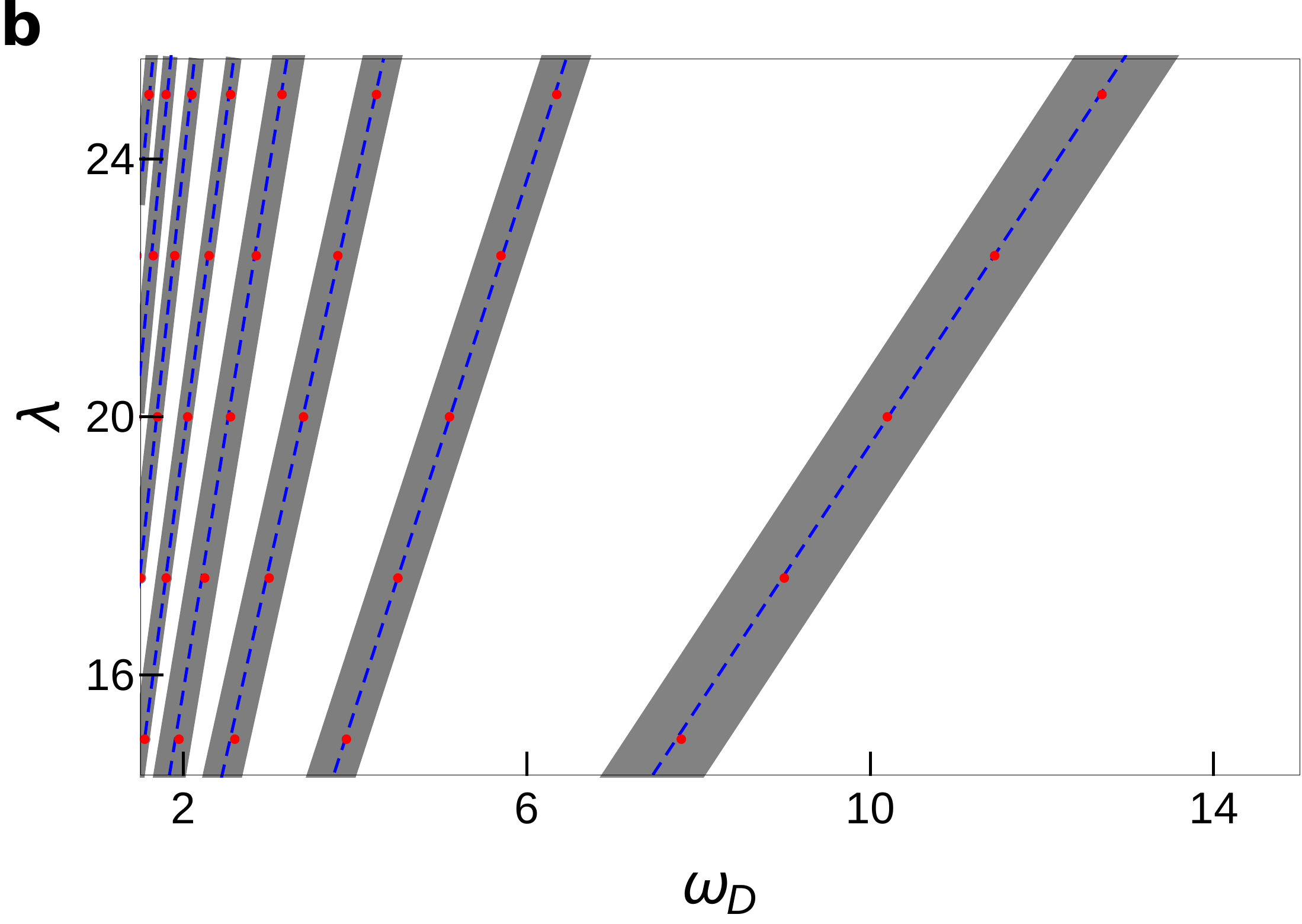}
\caption{Left Panel: Plots of half-chain entanglement entropy
$S_{L/2}$ for (a) $\hbar \omega_D= 8.25$  (b)$\hbar \omega_D= 8$,
(c) $\hbar \omega_D=7.75$, and (d) $\hbar \omega_D=7.5$. The plots
shows athermal scar states below the thermal band with low $S_{L/2}$
at $\hbar \omega_D=8.25$ and $\hbar \omega_D=7.5$ while no such
states are found for $\hbar \omega_D=8$ and $7.75$. The red circles
exhibits states which has high overlap with the initial $|{\mathbb
Z}_2\rangle$. For all plots, $\lambda=15$, $w=\sqrt{2}$ and $L=18$
and we have set $\hbar=1$. Right Panel: Phase diagram in the
$\lambda-\omega_D$ plane showing ergodic (shaded) and coherent
(white) regions with reentrant crossover between them. All other
parameters are same as those in the left panel. This figure is
adapted from Ref.\ \onlinecite{ryddyn1}.} \label{fig6}
\end{figure}

Naively, one would expect a crossover between these two regimes at
some intermediate drive frequency. In contrast, as shown, in the
right panel Fig.\ \ref{fig5}, the system exhibits non-monotonic
re-entrant behavior at multiple intermediate frequencies. This is
most easily noted by comparing the four plots in the right panel of
Fig.\ \ref{fig5}. Clearly, ETH predicted thermalization is restored
around $\hbar \omega_D \simeq 7.75 w$; however, scar induced
oscillations take over at a {\it lower} drive frequency $\hbar
\omega_D \simeq 7.5 w$. The corresponding half-chain entanglement
entropies, shown in the left panels of Fig.\ \ref{fig6}, indicate
the presence of scars at $\hbar \omega_D= 7.5 w$ and their absence
at $\hbar \omega_D=7.75 w$. Further studies, carried out in Ref.\
\onlinecite{ryddyn1}, demonstrates multiple crossovers between ETH
predicted ergodic and scar-induced coherent oscillatory behaviors at
intermediate frequencies leading to the phase diagram shown in the
right panel of Fig.\ \ref{fig6}. The density of the ergodic regimes
increases with lower frequency and they completely cover the phase
diagram at low drive frequencies leading to ergodic behavior in this
regime. However, at intermediate frequencies, as shown in Ref.\
\onlinecite{ryddyn1}, it is possible to tune into and out of such
ergodic regimes by tuning the drive frequency. This leads to the
possibility of drive-frequency induced tuning of ergodicity in a
driven non-integrable system.

This tunability of ergodicity can be qualitatively understood from
the analytical Floquet Hamiltonian (Eqs.\ \ref{forderfl} and
\ref{tordfl}) as follows. For $\hbar \omega_D \gg \lambda_0$, $\sin
\gamma \sim \gamma$ and $H_F^{(1)} \simeq H_{\rm PXP}$. Moreover
since $w/\lambda_0 \ll 1$, $H_F^{(3)}$ is negligible compared to
$H_F^{(1)}$ in this limit. Thus the Floquet Hamiltonian is of the
PXP form and supports quantum scars. For an initial state $|{\mathbb
Z}_2\rangle$, the dynamics gets maximal contribution from the scar
subspace leading to coherent oscillation \cite{scarlitrev}. In
contrast around $\gamma \simeq m \pi$ (where $m$ is a non-zero
integer), $H_F^{(1)} \to 0$ and $H_F^{(3)}$ dominates. The Floquet
Hamiltonian is then not of the PXP form anymore and does not support
scar states with large overlap with $|{\mathbb Z}_2\rangle$.
Consequently, the dynamics becomes ergodic around
$\omega_D=\omega_D^c = \gamma_0/(2 m \hbar)$; the width of the
ergodic regime depends on the relative magnitudes of $H_F^{(1)}$ and
$H_F^{(3)}$ as we move away $\omega_D^c$. Of course, this estimate
of $\omega_D^c$ does not take into account the renormalization of
$H_F^{(1)}$ from the higher order terms (such as the one from
$H_F^{(3)})$ and is thus not exact. However since these terms are
typically small in the intermediate frequency regime by a factor of
at least $w^3/\lambda_0^3 \ll 1$, this estimate turns out to be
qualitatively correct.

\begin{figure}[ht]
\includegraphics[width=0.49\linewidth]{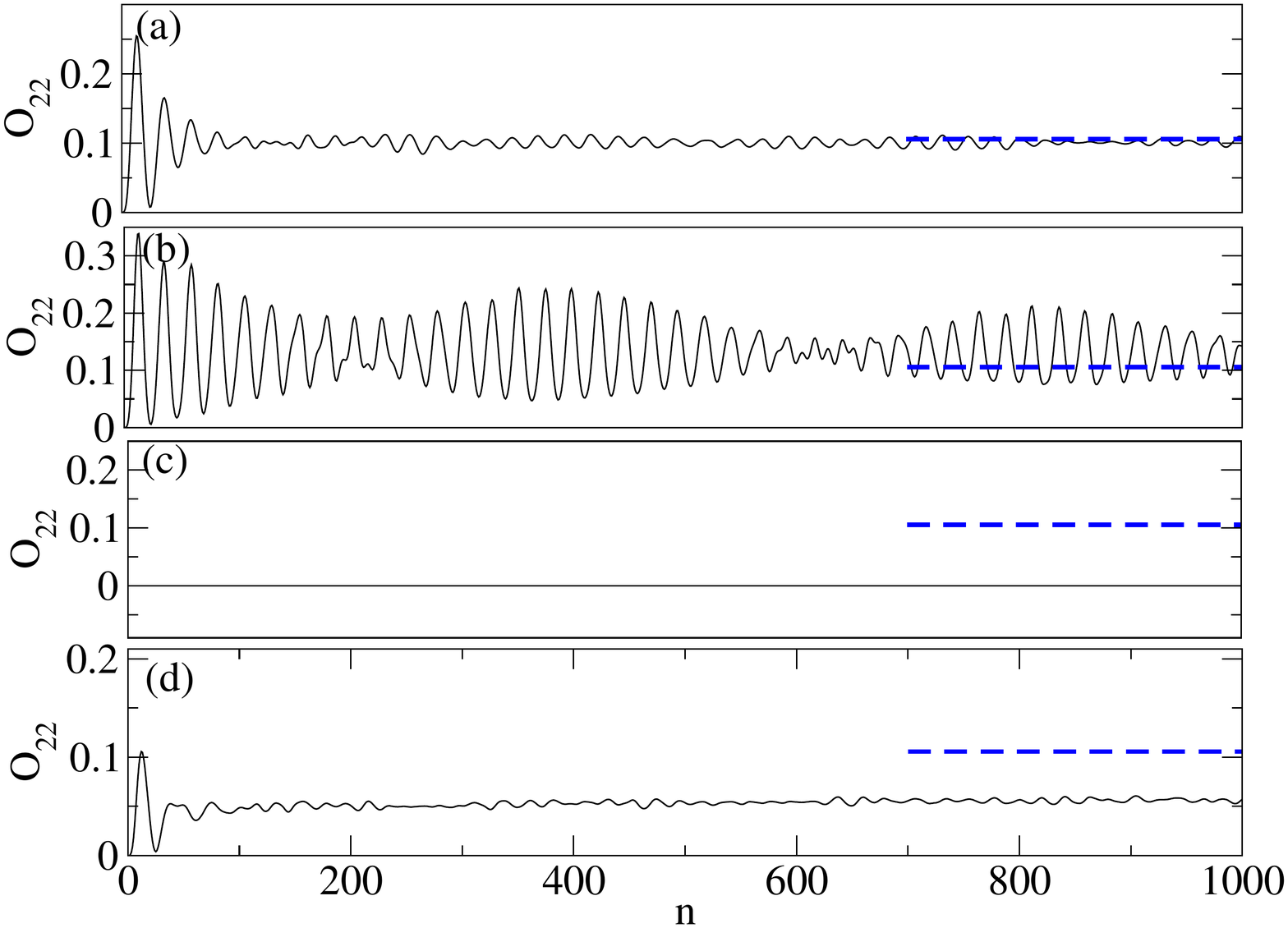}
\includegraphics[width=0.49\linewidth]{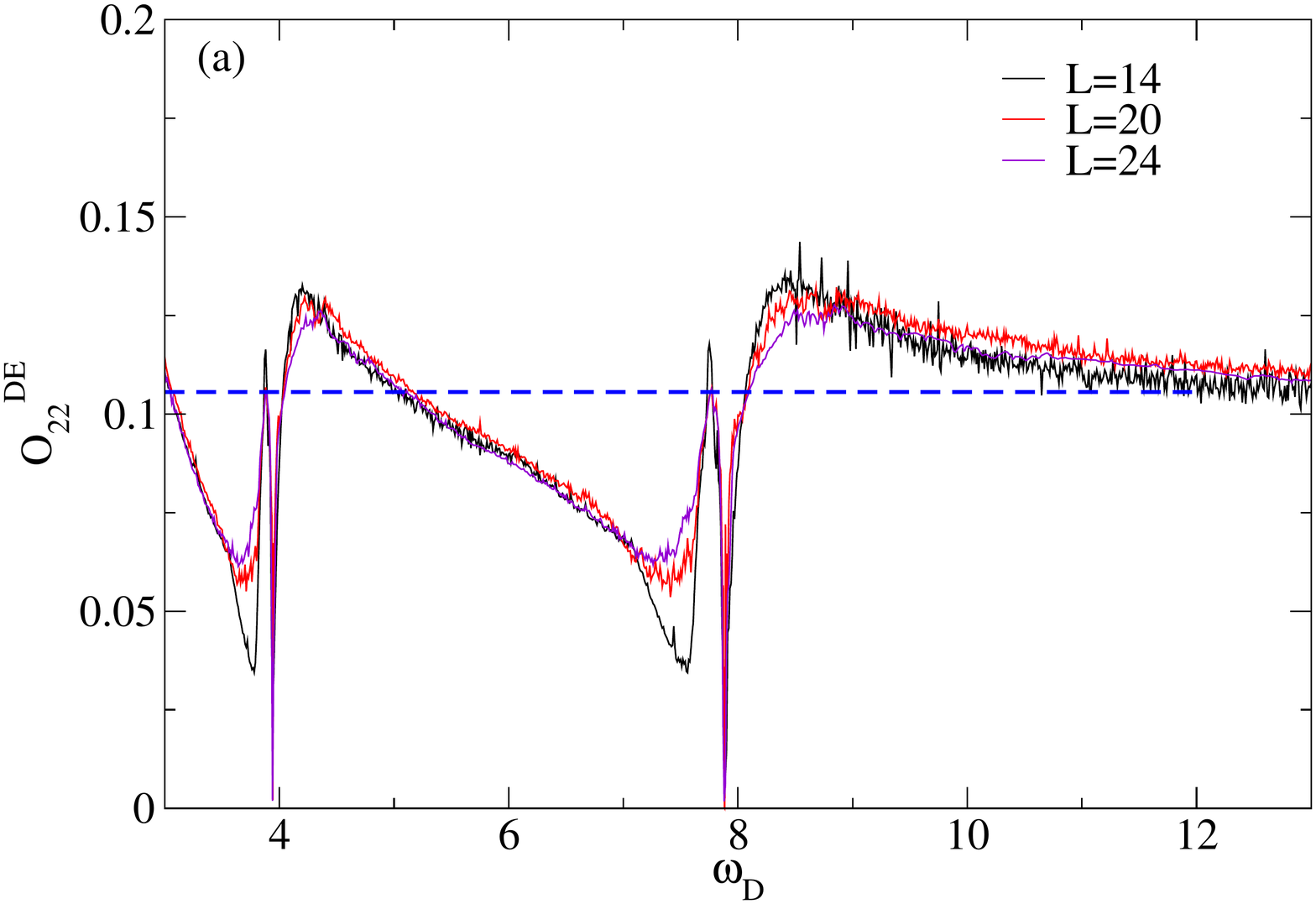}
\caption{Left Panel: Plot of $\langle O_{22}\rangle \sim O_{22}$ as
a function of $n$ for (a) $\hbar \omega_D= 100$ (b)$\hbar \omega_D=
8.5$, (c) $\hbar \omega_D=7.88$, and (d) $\hbar \omega_D=7.25$. For
all plots, $\lambda=15$, $w=\sqrt{2}$ and $L=26$ and we have set
$\hbar=1$. Right Panel: The steady state value of $\langle
O_{22}\rangle \equiv O_{22}^{\rm DE}$ as a function of $\omega_D$
for several $L$. For all plots $\lambda=15$, $w=\sqrt{2}$ and
$\hbar$ is set to unity. This figure is adapted from Ref.\
\onlinecite{ryddyn3}.} \label{fig7}
\end{figure}

Next, we consider the stroboscopic dynamics of $\langle
O_{22}\rangle$ starting from the $|0\rangle$ state. We note that for
$\hbar \omega_D /\lambda_0 \gg 1$, the dynamics exhibits
thermalization consistent with the ETH prediction as can be seen
from the left panel of Fig.\ \ref{fig5}. Thus in the quench limit,
scars do not play a role in the dynamics since they have negligible
overlap with the $|0\rangle$ state. As the drive frequency is
lowered, however, the situation changes as can be seen from left
panel of Fig.\ \ref{fig7}. Around $\hbar \omega_D= 8.5 w$, we find
the presence of long-time coherent oscillations while at $\hbar
\omega_D= 7.88 w$, $\langle O_{22} \rangle$  remains completely
frozen to its initial value. The latter behavior constitutes dynamic
freezing in an otherwise ergodic non-integrable system. Finally, for
$\hbar \omega_D=7.26 w$, we find that $\langle O_{22} \rangle$
reaches a steady state value which is lower than the ETH predicted
value (shown by the blue dotted line). This constitutes a
qualitatively different violation of ETH since coherent scar-induced
oscillatory dynamics (such as the one seen for $\hbar \omega_D=8.5
w$) usually leads to super-thermal steady state values in contrast
to the subthermal value found in the present case. The steady state
behavior of $\langle O_{22} \rangle $ starting from $|0\rangle$
initial state is shown in the right panel of Fig.\ \ref{fig7}. We
find that the steady state reaches its thermal value, shown by the
dotted line, at large $\omega_D$. As the frequency is lowered, it
reacher superthermal steady state value. Here the dynamics indicates
coherent long-time stroboscopic oscillation. Just below
$\hbar\omega_D <8 w$, the steady state value of $\langle O_{22}
\rangle$ drops and reaches zero at $\hbar \omega_D\simeq 7.88 w$.
This constitutes an example of dynamical freezing
\cite{dynfr1,dynfr2}. This is followed by a wide range of frequency
at which the steady state value remains subthermal. We also note the
presence of a second freezing point at around $\hbar \omega_D \simeq
3.95 w$. These steady state values of $O_{22}$ seem to be
independent of system size within the range of $L$ accessible within
ED.

\begin{figure}[ht]
\includegraphics[width=0.24\linewidth]{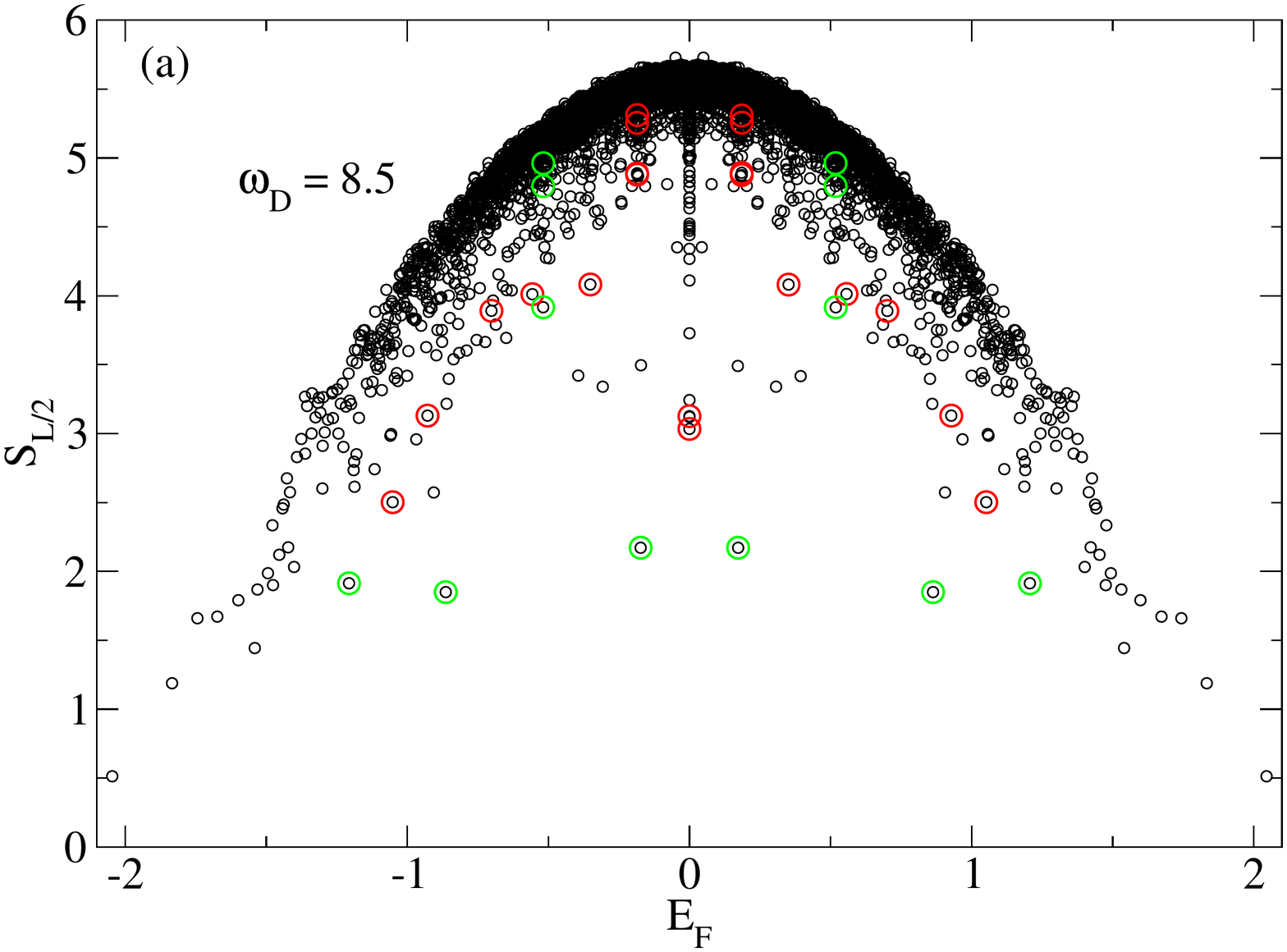}
\includegraphics[width=0.24\linewidth]{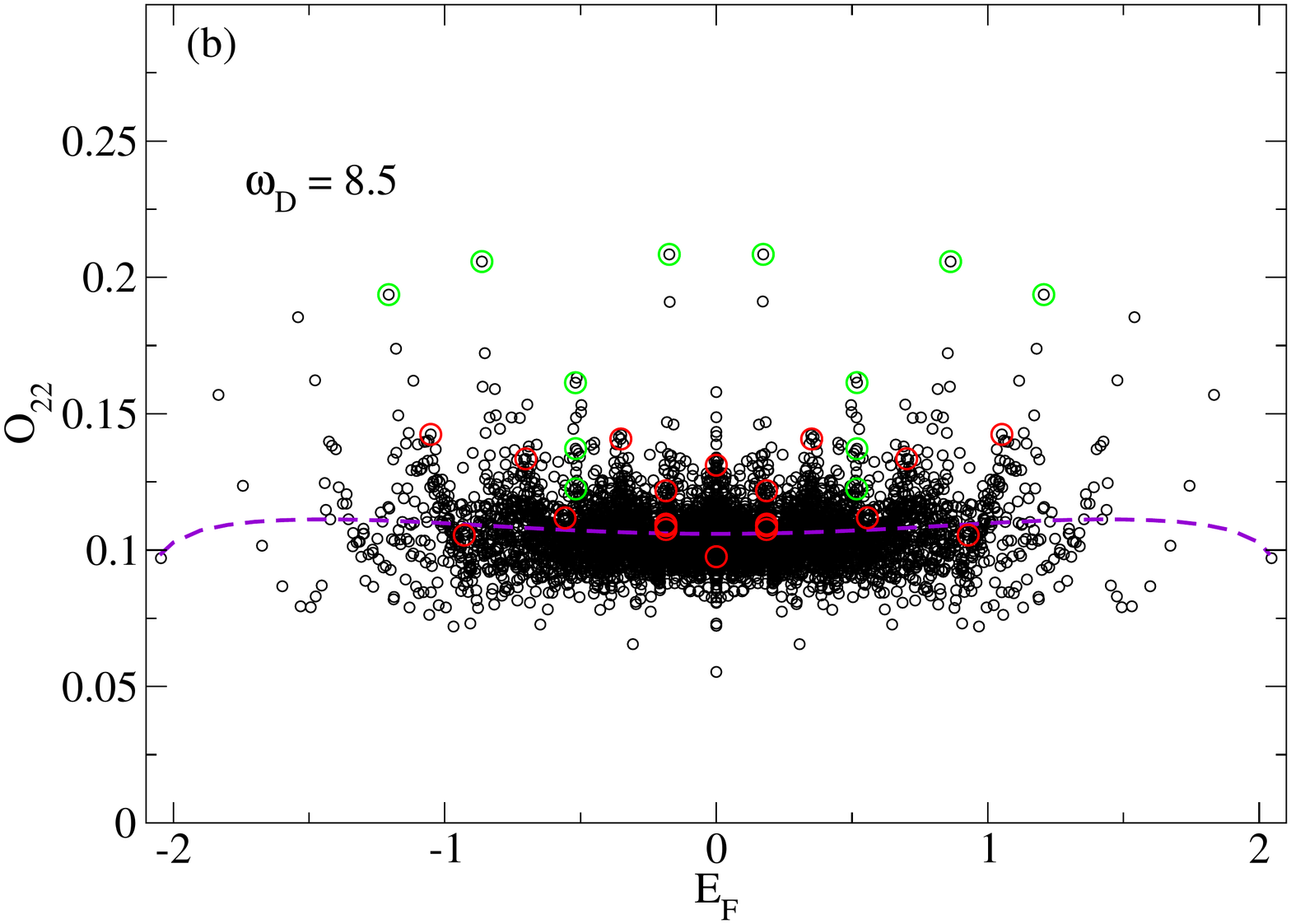}
\includegraphics[width=0.24\linewidth]{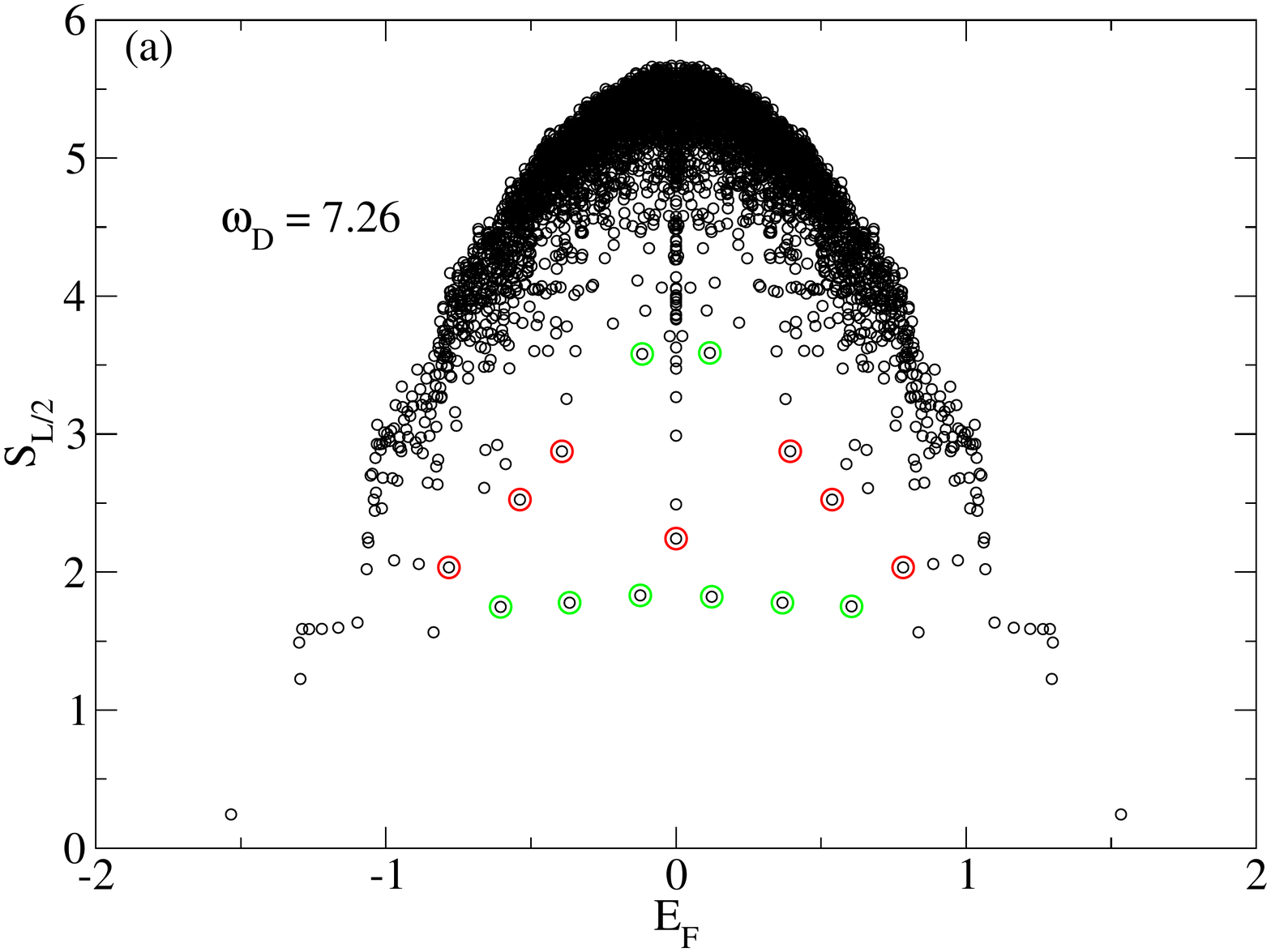}
\includegraphics[width=0.24\linewidth]{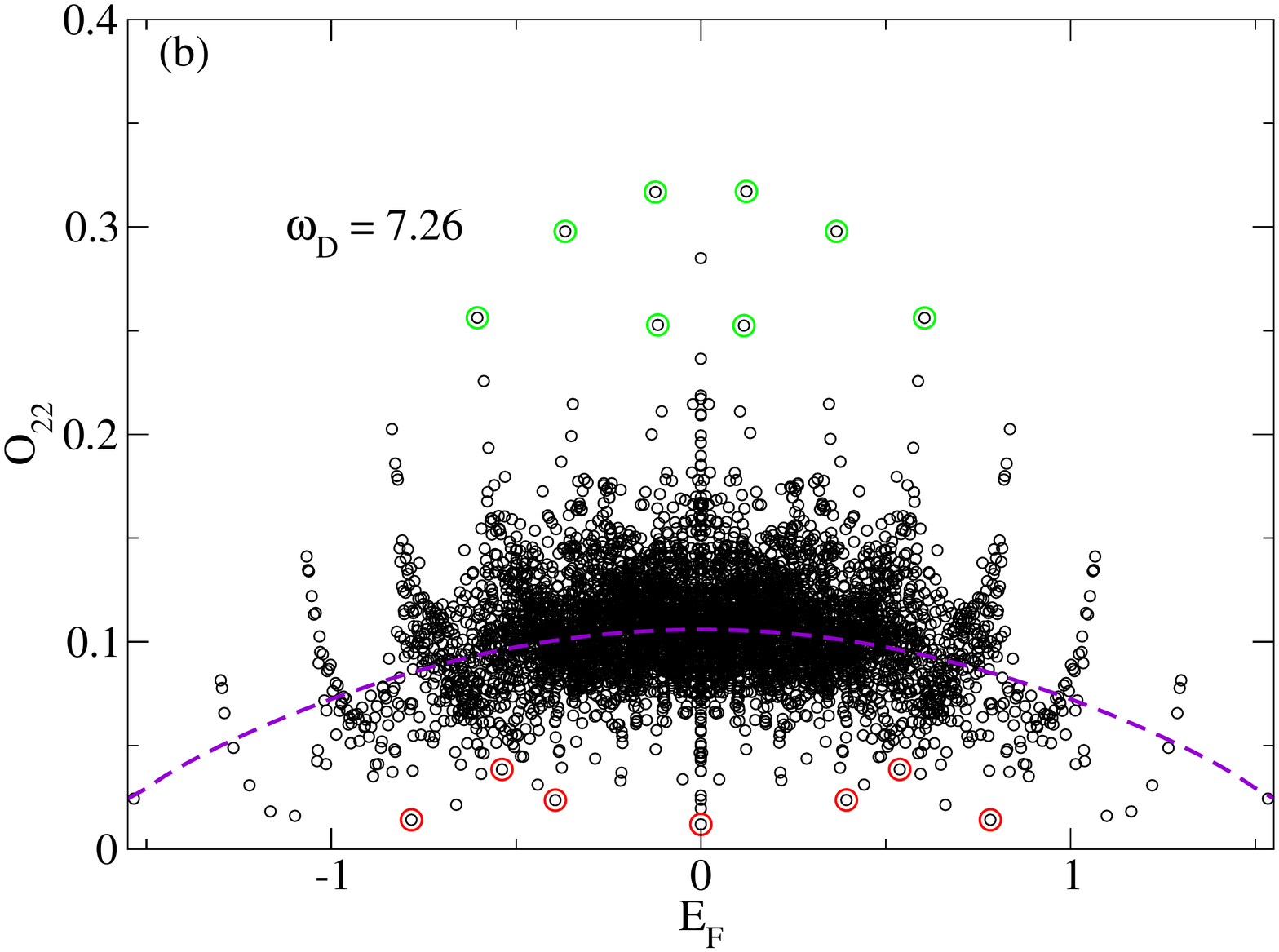}
\caption{Leftmost Panel: Plot of $S_{L/2}$ as a function of $E_F$
for $\hbar \omega_D= 8.5$. The green circles indicate states having
large overlap with $|{\bar Z}_2\rangle$ while the red circles
indicates those having large overlap with $|0\rangle$. For all
plots, $\lambda=15$, $w=\sqrt{2}$ and $L=26$. Left-center Panel:
Plot of $\langle|O_{22}|\rangle \equiv O_{22}$ for all Floquet
eigenstates. The red and the green circles indicates contribution
from states with large overlap with $|0\rangle$ and $|{\mathbb
Z}_2\rangle$ respectively. The violet dotted line indicates the ETH
prediction for $O_{22}$. Right-center and Rightmost Panels: Same as
the leftmost and the left-center panels respectively but for $\hbar
\omega_D=7.26$. This figure is adapted from Ref.\
\onlinecite{ryddyn3}.} \label{fig8}
\end{figure}

The oscillatory behavior of $O_{22}$ starting from $|0\rangle$ can
be understood by observing the nature of the Floquet eigenstates as
shown in the leftmost panel of Fig.\ \ref{fig8}. The plot shows the
presence of a different set of athermal states (indicated by red
circles) which has large overlap with the $|0\rangle$. These states
are different from the scars which has large overlap with $|{\mathbb
Z}_2\rangle$ shown by green circles. Thus the Floquet hamiltonian at
these frequencies supports at least two separate set of scar states;
this phenomenon has no analog in scars of the PXP model. Indeed, as
the frequency is increased, the set of scar states which has large
overlap with $|0\rangle$ merge into the continuum of thermal states
leaving behind only athermal $|{\mathbb Z}_2\rangle$ scars. This
brings out the central role of higher order terms in the Floquet
Hamiltonian for generation of such scars. The presence of such scars
lead to coherent oscillatory dynamics of $O_{22}$. Its steady state
value turns out to be superthermal which can be seen from the
left-center panel of Fig.\ \ref{fig8}. This is due to the fact that
for most scar eigenstates that have large overlap with $|0\rangle$
(shown as red circles in the left-center panel of Fig.\ \ref{fig8}),
$\langle O_{22}\rangle$ is larger than the ETH predicted thermal
value.

Similar scar states with large overlap with $|0\rangle$ is shown by
red circles in the right-center panel of Fig.\ \ref{fig8} for $\hbar
\omega_D= 7.26 w$. Here in contrast to the case $\hbar \omega_D=8.5
$, the initial state $|0\rangle$ turns out to have relatively large
overlap ($>0.01$) with a few scars states as shown in the figure and
small overlap with a large number of thermal states. The presence of
these thermal states do not allow for long time coherent
oscillations. However, the steady state value of $\langle O_{22}
\rangle$ is controlled by the athermal scar states since they have
relatively large overlap with $|0\rangle$. It turns out that these
athermal scar states have a subthermal value of $\langle
O_{22}\rangle$ (as can be seen from the rightmost panel of Fig.\
\ref{fig8}). Thus $\langle O_{22}\rangle$ quickly decays to a
subthermal steady state value leading to violation of ETH. We note
that such a violation do not involve oscillatory dynamics and thus
constitute a separate route of scar-induced ETH violation in finite
boson chains which has no analogue in chains subjected to quench.
\cite{ryddyn3}.

Finally, we discuss the phenomena of dynamic freezing at $\hbar
\omega_D \simeq 7.88$ for $w=\sqrt{2}$ and $\lambda=15$.
Qualitatively, the existence of such freezing is easy to understand
by noting that the state $|0\rangle$ is annihilated by all non-PXP
higher order terms of the Floquet Hamiltonian. This follows from the
fact that such terms necessarily involves $\tilde \sigma_j^-$
operator which annihilates the state $|0\rangle$. Thus for drive
period, where the coefficient of $H_{F}^{(1)}$ vanishes, $U(T,0) \to
I$ and one encounters freezing. This argument, in conjunction with
the first order perturbative Floquet Hamiltonian, suggest that the
freezing would occur at frequencies for which $w_r=0$: $\hbar
\omega_D= \lambda/(2n)$. This predicts a freezing frequency of
$\hbar \omega_D=7.5$ which differs quite a bit with the exact
result.

This discrepancy can be understood by noting that the first order
Floquet Hamiltonian neglects the normalization of $w_r$ due to
contribution of higher order term such as the one from $H_F^{(3)}$
(Eq.\ \ref{tordfl}). To see this, we consider an exact analytical
calculation of $H_F$ for $L=3$. Here the exact analytic calculation
is feasible since the Hilbert space consists of just two states in
the zero momentum sector; this reduces the problem to a driven
two-state problem \cite{ryddyn3}. These two states are $|0\rangle =
|\downarrow, \downarrow, \downarrow\rangle$ and $|1\rangle =
(|\downarrow, \uparrow, \downarrow\rangle + |\uparrow, \downarrow,
\downarrow\rangle + |\downarrow, \downarrow,
\uparrow\rangle)/\sqrt{3}$. The Hamiltonian, the space of these two
states, reads
\begin{eqnarray}
H_2(t) &=& \left( \begin{array}{cc} 0 & -\sqrt{3} w \\ -\sqrt{3} w &
\lambda(t)
\end{array} \right) \label{twoham}
\end{eqnarray}
For the square pulse protocol given by Eq.\ \ref{prot} the Floquet
Hamiltonian $H'_F$ corresponding to $H_2(t)$ can be exactly computed
\cite{ryddyn3}. A straightforward calculation shows that
\begin{eqnarray}
\langle 0| H'_F T |1\rangle &=& \cos^{-1} \left[1- \frac{12
w^2}{\Delta_0^2} \{1-\cos(\Delta_0 T/2) \}\right], \quad  \Delta_0=
\sqrt{ \lambda_0^2+ 12 w^2} \label{freezeqf}
\end{eqnarray}
This matrix element, which needs to be finite for the driven system
to evolve, vanishes for $\Delta_0 T= 4 m \pi$ (where $m$ is an
integer). For $\lambda=15$ and $w=\sqrt{2}$, $\hbar \omega_D \simeq
7.86$ for $m=1$, and this provides a near exact match to the
observed freezing frequency. Moreover, it also predicts the second
freezing point at $\hbar \omega_D \simeq 3.93$ which correspond to
$m=2$. We note that for $\lambda \gg w$, this frequency will reduce
to that predicted by first order Floquet theory ($\hbar \omega_D=
\lambda/2m$) as expected. The reason for the accuracy of the result
obtained with this simplistic calculation is that the higher order
multiple spin terms in $H_F$ do not contribute to the freezing
phenomenon as discussed earlier. We note that at the freezing point
the state $|0\rangle$ is disconnected from all other states in the
Hilbert space of the system; this constitutes an example of (weak)
fragmentation of the Hilbert space of $H_F$.

Before ending this section, we note that the periodic dynamics of
Rydberg atoms has been studied experimentally in Ref.\
\onlinecite{gr4}. The experiment used a continuous drive protocol
using a damped cosine drive and starting from $|Z_2\rangle$; it was
found the system exhibits a robust subharmonic response to the
drive. It was noted that this response depended on the initial state
and its relation to quantum scars in the system was discussed.
However, the dynamics of the system starting from $|0\rangle$ state
or that in the presence of a square pulse protocol has not been
studied in this work.

\section{Hilbert space fragmentation: A minimal model}
\label{fut}

It has been recently been pointed out that the presence of dynamical
constraints in a many-body system may lead to fragmentation of its
Hilbert space into several disconnected sectors. This phenomenon,
termed as Hilbert space fragmentation (HSF), provides yet another
route to violation of ETH in non-integrable quantum systems
\cite{fraglitrev, fraglit1,fraglit2,fraglit3,fraglit4,fraglit5,
fraglit6,fraglit7,fraglit8,bm1}. This phenomenon naturally arises in
constrained systems where the presence of additional conservation
laws provide the constraint
\cite{fraglitrev,fraglit1,fraglit2,fraglit3,fraglit4,fraglit5,fraglit6,fraglit7,fraglit8}.
The resulting physics has close similarity with those of fractons
\cite{fracrev1,fracrev2}. The realization of such constrained
systems using circuit models whose Hilbert space is identical to
that of a $S=1$ spin chain of length $L$ provides a neat example in
this context \cite{fraglit2}. In this case, it was shown that the
presence of two simultaneous $U(1)$ conserved quantities, namely
$Q=\sum_j S_j^z$ (which is analogous to total charge in a fractonic
model) and $P=\sum_j j S_j^z$ (dipole moment of a fractonic model),
provides the necessary constraints for fragmentation. Typically, in
most of the models studied, such conservation comes from the
commutation $[Q,H]=[P,H]=0$ and remains valid for all states in the
Hilbert space; moreover, they lead to an exponentially large number
of inert zero-energy states in the Hilbert space. We note here such
fragmentation may even separate states with same symmetry into
different, disconnected, segments within the Hilbert space. Such
fragmentation is dubbed as "strong" if the Hilbert space is
fragmented in exponentially many separate sectors; in this case, the
number of states which lead to violation of ETH increases
exponentially with system size \cite{fraglitrev,fraglit3}. If this
condition is not satisfied, the fragmentation is termed as weak; in
this sense, scars constitutes an example of weak fragmentation of
Hilbert space. A recent experiment involving Fermi-Hubbard model has
observed non-ergodic behavior in a titled Fermi-Hubbard system which
can possibly be attributed to such fragmemntation \cite{fragexp}. In
the rest of this section, we shall discuss such fragmentation in a
model which was inspired by the Floquet Hamiltonian of the tilted
Bose-Hubbard model; the details of HSF and its realization in other
model can be found in Ref.\ \onlinecite{fraglitrev}.

\subsection{Model and fragmentation}
\label{frag}

The tilted Bose-Hubbard model constitutes an example of a spin-half
system in a constrained Hilbert space. However, the model does not
show HSF. The reason for this becomes clear when one analyzes the
connectivity of the states of the model in the number basis in the
PXP limit. It turns out, as noted in Ref.\ \onlinecite{bm1}, that
most states in the Hilbert space are connected under the action of
$H_{\rm PXP}$ via the state $|0\rangle$. Thus a Hamiltonian which
would annihilate the state $|0\rangle$ may lead to HSF. Taking cue
from the structure of the Floquet Hamiltonian for the periodic
driven tilted Bose-Hubbard model, Ref.\ \onlinecite{bm1} pointed out
that one such possible model involving spin-half Pauli matrices on a
1D chain is given by
\begin{eqnarray}
H_{\rm fr} &=& w \sum_{j=1}^L \tilde (\sigma_{j-1}^+ \tilde
\sigma_{j+1}^+ \tilde \sigma_j^{-} +{\rm h.c.}), \label{fragham}
\end{eqnarray}
where $w$ is an arbitrary energy scale which shall be set to unity
for the rest of this section and it is understood that $H_{\rm fr}$
acts on the constrained Hilbert where two up-spins can not be
neighbors. We note that $H_{\rm fr}$ is identical to the three-spin
term in the third order Floquet Hamiltonian (Eq.\ \ref{tordfl}) but
with its coefficient set to unity. Such a term becomes the largest
term in $H_F$ around the point where renormalized $H_F^{(1)}$ (Eq.\
\ref{forderfl}) vanishes. We note that $H_{\rm fr}$ do not have any
simultaneous conserved quantities and thus differ from a class of
earlier studied models \cite{fraglitrev}.

The simplest class of states which demonstrates such fragmentation
corresponds to blocks of length $\ell=3$ with one or two up-spins in
a background of down-spins \cite{bm1}. These states can be written
as
\begin{eqnarray}
|X_{1,j}\rangle &=& |... 1_{j-1} 0_j, 1_{j-1} ...\rangle, \quad
|X_{2,j}\rangle = |... 0_{j-1} 1_j, 0_{j-1} ...\rangle
\label{hsfstates1}
\end{eqnarray}
where we have denoted up- and down-spins by $1$ and $0$ respectively
and ellipsis indicates down spins on all other sites. It is easy to
see that under action of $H_{\rm fr}$, these blocks transform to
each other: $H_{\rm fr}|X_{1,j}\rangle = |X_{2,j}\rangle$ and
$H_{\rm fr} |X_{2,j}\rangle = |X_{1,j}\rangle$. They are connected
to any other states in the Hilbert state and thus constitute a
fragment. Their linear combination
\begin{eqnarray}
|X_{\pm, j} \rangle &=& \frac{1}{\sqrt{2}} (|X_{1,j}\rangle \pm
|X_{2,j}\rangle) \label{eigenstates}
\end{eqnarray}
yields eigenstates of $H_{\rm fr}$ with eigenvalues $\pm 1$ (we have
set $w=1$). Moreover, if two such blocks are spaced with at least
two down-spins separating them, they act as non-interacting entities
and the total energy of the system becomes a sum of the energy of
the individual blocks. This leads to a class of eigenstates with
zero and integer (in units of $w$) energies: $E = (n_+-n_-)$, where
$n_{\pm}$ are the number of isolated $|X_{\pm}\rangle$ blocks in the
state. We note that these blocks are localized and are dubbed as
"bubbles" in Ref.\ \onlinecite{bm1}.

These bubbles develop dispersion when two individual bubbles are
allowed to interact by placing them next to one another. The
simplest of these states can be understood analytically, and they
form a closed Hilbert space fragment spanned by the states
\cite{bm1}
\begin{eqnarray}
|\psi_{1,k}\rangle &=&  \sum_j e^{i k j} T_j |... X_{2, \,j}
X_{2,\,j+3} ...\rangle \equiv \sum_j e^{i k j} T_j |... 0_{j-1} 1_j
0_{j+1}
0_{j+2} 1_{j+3} 0_{j+4} .... \rangle  \nonumber\\
|\psi_{2,k}\rangle &=&  \sum_j e^{i k j} T_j |... X_{1, \,j} X_{2,
\,j+3} ...\rangle \equiv \sum_j e^{i k j} T_j |... 1_{j-1} 0_j
1_{j+1} 0_{j+2} 1_{j+3} 0_{j+4} .... \rangle  \label{disstates}
\end{eqnarray}
where $T_j$ denotes the translation operator, the lattice spacing
has been set to unity,  and the ellipsis denotes all down spins. It
was shown in Ref.\ \onlinecite{bm1} that $|\psi_{1(2), k}\rangle$
form a closed subspace with
\begin{eqnarray}
H_{\rm fr} \left( \begin{array}{c} |\psi_{1,k}\rangle \\
|\psi_{2,k}\rangle \end{array} \right)  &=& \left( \begin{array}
{cc} 0 & (1 + e^{i k}) \\ (1+ e^{-i k}) & 0 \end{array} \right)
\left(
\begin{array}{c} |\psi_{1,k}\rangle \\ |\psi_{2,k}\rangle \end{array}
\right) \label{hamdis}
\end{eqnarray}
This leads to a pair of eigenstates in the momentum space given by
\begin{eqnarray}
E_k &=& \pm 2 \cos k/2   \label{enmom}
\end{eqnarray}
It was noted in Ref.\ \onlinecite{bm1} that Eq.\ ref{enmom} provide
an analytical explanation for the presence another class of
eigenstates with $E=\pm 2$ (for $k=0$), $E=\pm 1$ (for $k=2\pi/3$),
and $E=0$ (for $k=\pi$). Further for $L\ge 8$ such that $L=4n$ ($n
\in Z$), $k=\pi/2$ leads to $E= \pm \sqrt{2}$; this provides a
natural explanation of such eigenstates with simple irrational
eigenvalues that was found in the spectrum of $H_{\rm fr}$.

Apart from such simple fragments., more complicated fragments with
much longer bubbles leading to flat bands were discussed in Ref.\
\onlinecite{bm1}. We shall not discuss them in details here.
However, we would like to point out that the model exhibits a
phenomenon, dubbed as secondary fragmentation in Ref.\
\onlinecite{bm1}, which was not found in earlier works. Such
secondary fragmentation happens when basis states within a primary
fragments forms a further closed subspace under action of $H_{\rm
fr}$; these states are constructed out of specific linear
combination of a fixed number of basis states and they turn out to
be orthogonal to other states within the same primary fragment. The
eigenvalues corresponding to such eigenstates, for the model
discussed in Ref.\ \onlinecite{bm1}, are integers, and their number
increases with increasing $L$; importantly, their existence can not
be straightforwardly tied local classical conservation conditions.

\subsection{Adding a staggered field}
\label{stag}

In this section, we shall focus on the structure of the zero-energy
eigenstates of the model in the "bubble" sector. To this end, as
pointed out in Ref.\ \onlinecite{bm1}, it is useful to add a
staggered magnetic field to the model leading to
\begin{eqnarray}
H &=& H_{\rm fr} + H_{\Delta}, \quad H_{\Delta}= \frac{\Delta}{2}
\sum_j (-1)^j \sigma_j^z \label{stagf}
\end{eqnarray}
We note that there exists two operators
\begin{eqnarray}
Q &=& \prod_j \sigma_j^z, \quad C= \prod_j (\sigma_{2j}^z +
\sigma_{2j+1}^z)  \label{opeq}
\end{eqnarray}
We note that $[H,Q]=0$ so that the spectrum of $H$ is symmetric
around $E$. Moreover, we note that $H_{\Delta}$ anticommutes with
$C$. This allows for presence of zero energy modes of $H$. The
details of these zero modes has been worked out in Ref.\
\onlinecite{bm1}.

Since $H_{\Delta}$ is diagonal in the number basis, it makes sense
to work in the diagonal basis in $|X_{1(2)}\rangle$ states. In the
space of these states, it is possible to represent $H$ in terms of
pseudospin operators $\tau_j^{\alpha}$ ($\alpha=x,y,z$) such that
$\tau_j^x |X_{1,j}\rangle = |X_{2,j}\rangle$. In terms of these
pseudospin operators, one can write $H$ as \cite{bm1}
\begin{eqnarray}
H &=& \epsilon_0 \sum_j \sum_{\alpha=x,y,z} \eta^{\alpha} \tilde
\tau_j^{\alpha} +\frac{\Delta}{2} (n_{\rm odd}-n_{\rm even}), \quad
\epsilon_0 =\sqrt{1+9 \Delta^2/4}, \quad \eta = (\eta^x, \eta^y,
\eta^z)= (1, 0, 3\Delta/2)/\epsilon_0 \label{heffbubble}
\end{eqnarray}
where $n_{\rm odd/even}$ are the number of bubbles centered on odd
or even sites. Here we have defined $\tilde \tau^{y,z}_j = (-1)^j
\tau^{y,z}_j$ and $\tilde \tau^x_j=\tau_j^x$ for all $j$, where $j$
denotes the center of the bubbles and the sum over $j$ indicates sum
over number of such bubbles. Thus we find $H$ reduces to a
collection of non-interacting pseudospins (with $s=1/2$) on every
site. This indicates that the sector will contribute states to the
Hilbert space of $H$ which has area-law entanglement. Moreover,
these states indicates a novel features when it comes to
out-of-equilibrium dynamics of initial states belonging to the
bubble sector.

To see this, let us consider a square pulse protocol for which
$\Delta(t)= -(+)\Delta_0$ for $t \le (>) T/2$, where $T=2
\pi/\omega_D$ is the driving frequency. We shall assume that we
start the dynamics from a state which belongs to the bubble sector;
HSF then ensures that the dynamics will be controlled by states
within the sector. Since $H$ constitutes non-interacting pseudospins
on every center site of a bubble, $U(T,0)$ corresponding to a
square-pulse drive protocol can be found exactly. The reason for
this stems from the fact that here one deals with a $s=1/2$
pseudospin on every such site; this is similar to an analogous
problem for Ising or other integrable model where an analogous
structure can be seen in momentum space \cite{bookchap1}. A
straightforward analysis yields for $n_{\rm odd}=n_{\rm even}$
\cite{bm1}
\begin{eqnarray}
U(T,0) &=& \prod_{j=1.. L} \left( \begin{array} {cc} p_j & q_j \\
-q_j^{\ast} & p_j  \end{array} \right), \quad  p_j = \frac{(3
\Delta_0/2)^2+ \cos(\epsilon_0 T)}{\epsilon_0^2}, \quad q_j =\frac{
(-1)^j(3\Delta/2)[1-\cos(\epsilon_0 T)] - i \epsilon_0 \sin
(\epsilon_0 T)}{\epsilon_0^2} \label{ueq1}
\end{eqnarray}
This shows that $U(T,0) = I$ when $\omega_D= \omega_f =
\epsilon_0/n$. At these frequencies, the stroboscopic dynamics of
the system, starting from any of the states in the bubble sector,
exhibits dynamical freezing \cite{dynfr1,dynfr2,uma1,ryddyn3}. Such
freezing clearly stems from fragmentation and is a non-perturbative
exact phenomenon. Moreover, its existence shows up in the dynamics
of simple initial states in the Fock space (such as
$|X_{1,j}\rangle$) making it experimentally accessible.

\subsection{Connection to lattice gauge theories}
\label{lgt}

Finally, we point out the connection of these systems to lattice
gauge theories. The possibility of simulating lattice gauge theories
using optical lattice systems has been a subject of recent interest
\cite{lgt1,lgt1a,lgt2,lgt3,lgt4,iaed1}. The reason for this is
partly the possibility of realization of an experimental platform
for study of confinement. It is well-known from the seminal work of
Ref.\ \onlinecite{coleref} that in 1D quantum electrodynamics,
charges (charge $\pm e$ for particles and anti-particles) display
the phenomenon of confinement in the presence of a background of
electric field $E_b$. This confinement stems from the fact that the
energy of these charges increases linearly with the distance between
them. This is characterized by a parameter $\theta$ which is
proportional to the background electric field $E_b$: $\theta= 2 \pi
E_b/e$. It was shown for all $\theta \ne \pi$, a pair of charges
remain confined since their energy grows linearly with distance when
they are attempted to be separated. It was also pointed out in Ref.\
\onlinecite{coleref} that this form of confinement holds in $d=1$;
for higher dimensions, the presence of transverse photons changes
the scenario and leads to deconfinement.

A variant of this phenomenon is expected to be found in possible
realization of U(1) lattice gauge theories, generically termed as
quantum link models \cite{lgt1,lgt1a,lgt2,lgt3,lgt4,iaed1}, using
optical lattice platforms. Indeed, a recent work on the PXP model in
the presence of an additional staggered magnetic field of strength
$\Delta$ showed that such a model can be mapped to a $U(1)$ lattice
gauge theory with $\Delta = J(\theta/\pi -1)$ \cite{iaed1}, where
$J$ is a microscopic energy scale. Thus the absence of $\Delta$
which corresponds to $\theta=\pi$ in the gauge theory language,
corresponds to the PXP model which shows deconfined behavior. In
contrast, the presence of a large $\Delta$ leads to confining
behavior whose signature can be picked up in quench dynamics of such
system \cite{iaed1}. We note here that an active field of research
in this area involves understanding the role of gauge invariance in
the dynamical evolution of these systems both experimentally
\cite{gaugeexpt1} and theoretically \cite{gaugeth1,gaugeth2}.

To understand the mapping of $H_{\rm fr}$ to lattice gauge theory,
we begin by writing the spin variables in the language of
Kogut-Susskind fermions \cite{kosuss1}. This provides a simple
dictionary which relates the Rydberg spins $\sigma_j^{\alpha}$ to
the fermionic matter (denoted by fields $\psi$) and gauge fields
(denoted by field $\hat E$) which are the degrees of freedom in the
gauge theory. The gauge (electric) field and the Rydberg spins live
on the link $\ell$ of the dual 1D lattice ( {\it i.e.} sites of the
original lattice) while the fermionic matter fields live on the
sites $j$ of the dual lattice. The electric fields take value $\pm
1/2$ and are related to the Rydberg spins via the relation
\cite{bm1}
\begin{eqnarray}
E_{\ell} &=& S_{\ell}^z = \eta_{j} \sigma_{\ell}^z /2  \label{ef}
\end{eqnarray}
where $\eta_j= \pm 1$ refers to the site $j$ at the left of the link
$\ell$ and is $1(-1)$ if that site is odd(even). The fermionic
matter field $\psi_j$ is related to the electric field by the
Gauss's law $G_j=0$ where \cite{lgt1}
\begin{eqnarray}
G_j &=& E_{\ell} - E_{\ell +1} - \hat n_j - [1 +(-1)^j]/2, \quad
\hat n_j = \psi_j^{\dagger} \psi_j \label{glaw}
\end{eqnarray}
It turns out that the constraint of having no two up-spins as
neighbors is exactly implemented by this law; moreover, the number
of gauge-invariant states for a chain of length $L$ exactly equals
the number of states within the constrained Hilbert space of the PXP
model. In this language, the PXP model, supplemented by the
staggered magnetic field term, can be written as \cite{iaed1}
\begin{eqnarray}
H_{\rm spin} &=&  - \sum_j \tilde \sigma_j^x + \frac{\Delta}{2}
\sum_j (-1)^j \sigma_j^z  \equiv  - \sum_i (\psi_{i}^{\dagger}
U_{\ell} \psi_{i+1} +{\rm h.c.}) + m \sum_i \psi_i^{\dagger} \psi_i
\label{gaugepxp}
\end{eqnarray}
where $U_{\ell} = S_{\ell}^+$ leading to $[E_{\ell},U_{\ell}]=
U_{\ell}$ and $\ell$ joins the dual lattice sites $i$ and $i+1$.
Thus the staggered spin term acts as the mass of the fermion fields.

It turns out that $H_{\rm fr}$ can also be written in the gauge
theory language leading to a Hamiltonian \cite{bm1}
\begin{eqnarray}
H_{\rm fr} &=& \sum_{i=1}^L \psi_i^{\dagger} U_{\ell} U_{\ell +1}
U_{\ell+2} \psi_{i+3} \Gamma_{i+1} \Gamma_{i+2} +{\rm h.c.}
\label{lgtfrham}
\end{eqnarray}
where $\ell$ is the link between the sites $i$ and $i+1$ on the dual
lattice and $\Gamma_i = \hat n_i $ for odd and $(1-\hat n_i)$ for
even $i$ respectively. It was shown in Ref.\ \onlinecite{bm1} that
$[H_{\rm fr}, G_j]=0$ which ensures that $H_{\rm fr}$ obeys Gauss's
law. The key point about the gauge theory representations of $H_{\rm
fr}$ (Eq.\ \ref{lgtfrham}) is that the it only conserves local
charge. There is no dipole moment conservation associated with this
model unlike most models of Hilbert space fragmentation (see Refs.
\onlinecite{fraglit1,fraglit2,fraglit3,fraglit4}) studied earlier.

Instead, such fractonic behavior may appear as emergent phenomenon
in certain sectors of theory \cite{bm1}. This was demonstrated in
Ref.\ \onlinecite{bm1} in the bubble sector. To see this let us
consider a single bubble state. All the down spin outside this
bubble are annihilated by $H_{\rm fr}$. In the gauge theory
language, this means that within this sector, all the fermionic
charges outside the bubble are immobile. It was shown that the
dynamics involve only the bubble which, here, constitutes dipoles
involving three lattice sites (length three dipoles). Only these
dipoles have significant dynamics under action of $H_{\rm fr}$ and
this leads to emergent fractonic physics. We note that this
emergence occurs in a fragment of the Hilbert space which is not
necessarily a low-energy sector. Other fragmented sectors may show
similar emergence, and this is discussed in details in Ref.\
\onlinecite{bm1}.

\section{Discussion}
\label{conc}

In this review, we have touched upon several aspects of the tilted
boson chain which can be realized in 1D optical lattices hosting
trapped ultracold bosons. It turns out that the physics of this
system has close parallel to that of Rydberg atoms in the sense that
both systems are described by similar effective Hamiltonians in
their low-energy sector.

The ground state phase diagram of such bosonic system provides a
route to realizing translational symmetry broken Mott states. In
addition, they also host a quantum phase transition between Mott
states with broken and unbroken translational symmetries. For the
tilted boson chain, this transition belongs to the Ising
universality class and can be understood in terms of a dipole model
of the bosons. A modification of this dipole model may lead to
realization of non-Ising critical points, as has been seen in
Rydberg atom arrays.

The quench dynamics of these systems provided the key clue to
unraveling the presence of quantum scars in their Hilbert space.
Such states lead to a weak violation of the eigenstate
thermalization hypothesis in finite chain as has been seen in recent
experiments involving these systems. The ramp dynamics of these
systems near their critical point may provide a platform to test
Kibble-Zurek scaling law; this is particularly interesting for the
Rydberg arrays where the transition is non-Ising like.

The study of periodic dynamics of such systems shows the possibility
of tuning their ergodicity properties using the drive frequency. It
was found that these systems shows unconventional phenomenon such as
the presence of sub-thermal steady states, long term coherent
oscillations, and dynamical freezing in the presence of a periodic
drive. These effects can be analyzed using the Floquet Hamiltonian
of these driven system; moreover, they occur, in contrast to their
quench counterparts, for both $|0\rangle$ and $|{\mathbb
Z}_2\rangle$ initial states and constitute different routes to
violation of ETH in finite-sized chains.

The Floquet Hamiltonian of this periodically driven system provides
a class of terms which can act as a minimal model for $s=1/2$ spins
exhibiting HSF. This model, unlike some earlier models displaying
HSF, do not have simple conserved quantities which causes the
fragmentation; instead, such conservation emerges in specific sector
of the model. Also, the model provides an example of secondary
fragmentation which does not follow from classical conservation
laws.  As a consequence of HSF, the model exhibits exact dynamical
freezing for an infinite number of drive frequencies and for an
exponentially large number initial classical Fock states.

Several aspects of this tilted boson chains and/or Rydberg ladders
remain to be studied. These include the effect of staggered detuning
term on the driven system, the physics of multiple interacting
Rydberg chains and manifestation of the physics of these driven
system in higher dimensions. These studied are expected to add to
the already large class of the physical phenomenon seen in these
systems. \\ \\

{\it Acknowledgement}: KS thanks D. Banerjee, B. K. Clark, U.
Divakaran, P. Fendley, S. M. Girvin, R. Ghosh, S. Kar, M.
Kolodrubetz, B. Mukherjee, S. Nandy, D. Pekker, S. Powell, S.
Sachdev, A. Sen, and D. Sen for collaborations on several related
projects.

\end{document}